\documentclass[aps,prl,twocolumn,superscriptaddress,longbibliography]{revtex4-2}

\usepackage{amsmath,amssymb,amsfonts}
\usepackage{bm}
\usepackage{mathtools}
\usepackage{physics}
\usepackage{graphicx}
\usepackage{xcolor}
\usepackage{microtype}
\usepackage{algorithm}
\usepackage{algpseudocode}
\usepackage{hyperref}
\usepackage{orcidlink}
\usepackage{tikz}
\usepackage{overpic}
\usepackage{siunitx}
\hypersetup{colorlinks=true,linkcolor=blue,citecolor=blue,urlcolor=blue}

\newtheorem{theorem}{Theorem}

\algrenewcommand\algorithmicrequire{\textbf{Input:}}
\algrenewcommand\algorithmicensure{\textbf{Output:}}

\newcommand{\dl}{\delta\lambda}
\newcommand{\deltamin}{\Delta_{\rm min}}
\newcommand{\bth}{\bm{\theta}}
\newcommand{\nupdate}{N_{\rm updates}}
\newcommand{\bigo}{\mathcal{O}}
\newcommand{\bg}{\bm{g}}
\newcommand{\tref}[1]{Theorem~\ref{#1}}
\newcommand{\eref}[1]{Eq.~\ref{#1}}

\newcommand{\sm}{\cite{supp}}
\newcommand{\Var}{\mathrm{Var}}
\newcommand{\Std}{\sigma}

\makeatletter
\newcommand{\prlsection}[1]{%
  \par\addvspace{0.8ex}%
  \noindent{\normalfont\itshape\bfseries #1\unskip\enspace---\enspace}%
  \ignorespaces
}
\makeatother

\makeatletter
\renewenvironment{description}
  {\list{}{\labelwidth=0pt \leftmargin=0pt \itemindent=0pt
           }}
  {\endlist}

\makeatother

\newcommand{\orcBojan}{\orcidlink{0000-0002-5085-2506}}
\newcommand{\orcMarco}{\orcidlink{0000-0002-3215-3453}}
\newcommand{\orcLewis}{\orcidlink{0009-0006-1595-102X}}
\newcommand{\orcMichael}{\orcidlink{0000-0002-2636-9936}}

\newcommand{\affA}{\affiliation{Faculty of Computer and Information Science, University of Ljubljana, Ve\v{c}na pot 113, 1000 Ljubljana, Slovenia}}
\newcommand{\affB}{\affiliation{Rudolfovo, Science and Technology Center, Novo mesto, Slovenia}}
\newcommand{\affC}{\affiliation{Quantinuum, Partnership House, Carlisle Place, London SW1P 1BX, United Kingdom}}

\begin{document}

\title{Trainable, self-verifying variational quantum eigensolver using adiabatic warm starts}

\author{Bojan Žunkovič\orcBojan}\email{bojan.zunkovic@fri.uni-lj.si}\affA\affB
\author{Marco Ballarin\orcMarco}\affC
\author{Lewis Wright\orcLewis}\affC
\author{Michael Lubasch\orcMichael}\email{michael.lubasch@quantinuum.com}\affC

\date{October 29, 2026}

\begin{abstract}
We study an adiabatic variant of the variational quantum eigensolver (VQE) in which VQE is performed iteratively for a sequence of Hamiltonians along an adiabatic path.
We derive the conditions under which gradient-based optimization successfully prepares the adiabatic ground states.
These conditions define a trainable tracking region.
Additionally, when combined with a gap lower bound, energy-standard-deviation measurements certify eigenstate accuracy at run-time and verify convergence to the global optimum.
\end{abstract}

\maketitle

\prlsection{Introduction}
The variational quantum eigensolver (VQE) is a promising algorithm for ground-state preparation on current noisy intermediate-scale quantum (NISQ) computers~\cite{CeEtAl21, BhEtAl22, TiEtAl22}.
In its traditional formulation, however, VQE faces significant obstacles, including challenging optimization landscapes characterized by barren plateaus~\cite{LaEtAl25} and numerous local optima~\cite{AnKi22}, as well as the absence of an intrinsic mechanism to verify the correctness of the obtained solutions~\cite{GhKaKa19}.

An alternative to VQE, which can avoid these obstacles, is adiabatic ground-state preparation.
In its standard circuit-model implementation~\cite{FaEtAl00, AlLi18}, however, this approach often generates quantum circuits that are too deep to be feasible on NISQ devices.
Various NISQ-compatible schemes have been proposed --- variational path-following and Hamiltonian-ansatz-based methods~\cite{FaGoGu14, WeHaTr15, GaLa18, ChEtAl20, HaEtAl22, ScTuCi22, KoEtAl24, SaEtAl25}, digitized and counterdiabatic protocols~\cite{HeEtAl21, ZhEtAl21, ChEtAl22, HeChSo22, KaEtAl25, RoEtAl25, BaEtAl26}, and compression or meta-learning approaches~\cite{CeKoAs21, ChEtAl23, MiHsZh24, McLu24}.

Closely related to adiabatic methods is the concept of warm start, which has recently gained attention in variational quantum optimization~\cite{PuEtAl25, MhEtAl25, BaEtAl25, PuEtAl26}.
Several adiabatically inspired methods already reuse parameters along a path, whereas these recent works study warm-start initialization and landscape guarantees more generally.
The central idea is to optimize the variational ansatz incrementally along a predefined optimization path.
At each point along this path, the optimizer initializes the ansatz with the parameters from the previous point, thereby warm-starting the optimization.
It has been shown that warm-start strategies can provably avoid the barren plateau problem in variational quantum simulation~\cite{PuEtAl25}, multivariate state preparation~\cite{BaEtAl25}, and VQE optimization using adiabaticity~\cite{PuEtAl26}.

\begin{figure}
\centering
\includegraphics[width=0.9\linewidth]{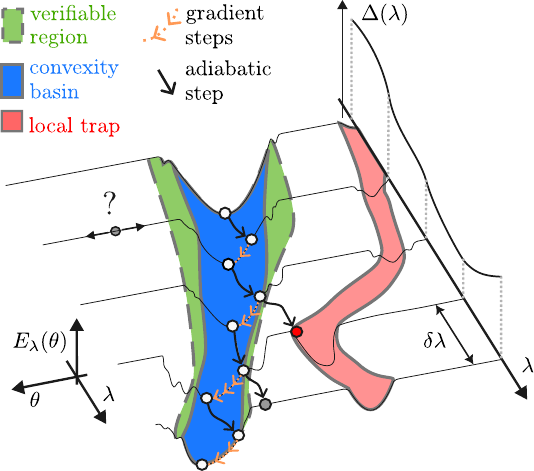}
\caption{\label{fig:1}
AVQE optimization landscape ---
Consider a Hamiltonian $H(\lambda)$ where $\lambda \in [0, 1]$ defines an adiabatic path, a variational ansatz parametrized by $\theta$, and the associated energy-based cost function $E_{\lambda}(\theta)$.
Starting from the ground state at $\lambda = 0$, there exists a basin (blue; Theorem~\ref{thm: tracking}) that follows the adiabatic path and guarantees convexity.
The size of the convexity basin decreases at most linearly with the instantaneous gap $\Delta(\lambda)$ (black curve, right-hand two-dimensional plot), which denotes the energy difference between the ground and first excited states at $\lambda$.
Outside the convexity basin, the optimization landscape may exhibit barren plateaus (white) and local minima (red), which can hinder gradient-based optimization.
In the verifiable region (green; Theorem~\ref{thm: certificate}), which also scales linearly with $\Delta(\lambda)$, we cannot assume convexity, but we can identify a unique nearby eigenstate.
Ground-state initialization and stepwise tracking identify it with the ground-state branch.
This enables adaptation of the adiabatic step width $\delta\lambda$ to ensure provable convergence to the target ground state at $\lambda = 1$.
}
\end{figure}

In this paper, we analyze the warm-start paradigm in an adiabatic VQE (AVQE), where VQE is executed sequentially along the Hamiltonians of a discretized adiabatic path.
We determine the conditions under which gradient descent tracks the instantaneous ground states, introduce a run-time verification procedure based on the energy standard deviation, provide gap-conditional correctness certificates, present numerical experiments, and analyze robustness against measurement shot noise.
Figure~\ref{fig:1} visualizes our key findings.
Our results establish that, under the stated gap, representability, and regularity assumptions, AVQE can be formulated as a trainable, self-verifying algorithm for ground-state computations compatible with NISQ-style circuits.

\prlsection{Setup}
We consider a linear adiabatic schedule defined by the Hamiltonian
\begin{equation}
H(\lambda) = (1-\lambda)H_\text{i} + \lambda H_\text{f}, \qquad \lambda\in[0,1],
\label{eq:adiabatic_hamiltonian}
\end{equation}
where the initial Hamiltonian $H_\text{i}$ is chosen to have an easy-to-prepare ground state and the final Hamiltonian $H_\text{f}$ encodes the desired target ground state.
We denote the instantaneous ground state by $\ket{\psi_0(\lambda)}$, the spectral gap by
\begin{equation}
\Delta(\lambda)=E_1(\lambda)-E_0(\lambda)
\end{equation}
where $E_0(\lambda)$ is the ground and $E_1(\lambda)$ the first-excited-state energy at $\lambda$, and we denote the minimum gap along the path by
\begin{equation}
\Delta_{\min}=\min_{\lambda\in[0,1]}\Delta(\lambda).
\end{equation}

We use the variational ansatz
\begin{equation}
\ket{\psi(\bm{\theta})}=U_{M-1}\cdots U_0\ket{0}, \qquad U_j=e^{-iP_j\theta_j},
\label{eq:ansatz}
\end{equation}
where bold symbols denote vectors, $P_j\in\{I,X,Y,Z\}^{\otimes n}$, and $M$ is the number of variational parameters.
For a general smooth ansatz, analogous conclusions follow with ansatz-specific smoothness and curvature bounds, but without the explicit dependence on the number of variational parameters.

The variational energy at fixed $\lambda$ is
\begin{equation}
E_\lambda(\bth) = \bra{\psi(\bth)}H(\lambda)\ket{\psi(\bth)},
\end{equation}
and we denote by $\bth^\star(\lambda)$ a global minimizer satisfying \(\ket{\psi(\bth^\star(\lambda))}=\ket{\psi_0(\lambda)}\).

The variational manifold is equipped with the quantum metric, defined as the real part of the quantum geometric tensor:
\begin{equation}
g_{\mu\nu}(\bth(\lambda)) = \mathrm{Re}\!\left( \langle \partial_\mu \psi | \partial_\nu \psi \rangle - \langle \partial_\mu \psi | \psi \rangle \langle \psi | \partial_\nu \psi \rangle \right),
\label{eq:metric}
\end{equation}
which determines the local behavior of the variational optimization landscape.

We discretize the adiabatic parameter into $T$ slices,
\begin{equation}
\lambda_t = t \, \delta \lambda, \qquad \delta \lambda = \frac{1}{T}, \qquad t=0,1,\dots,T,
\end{equation}
and perform optimization sequentially along the path using warm starts.
We initialize $\bth_{0}=\bth^\star(0)$.
For each slice $t=1,\dots,T$, we run $K$ steps of gradient-based optimization on $E_{\lambda_t}$, starting from the previous parameter values $\bth_{t-1}$:
\begin{align}
\bth_{t}^{(0)} &= \bth_{t-1}^{(K)}, \\
\bth_{t}^{(k+1)} &= \bth_{t}^{(k)} - \eta\, \bm{\mathcal{G}}_{t}\!\left(\bth_{t}^{(k)}\right), \quad k=0,1,\dots,K-1, \label{eq:warmstart_update}\\
\ket{\psi_t} &= \ket{\psi(\bth_t^{(K)})},
\end{align}
where $\eta>0$ is the learning rate and $-\bm{\mathcal{G}}_t(\bth)$ denotes the chosen descent direction for the objective at $\lambda=\lambda_t$.
In the simplest case, $\bm{\mathcal{G}}_t(\bth)=\nabla_{\bth} E_{\lambda_t}(\bth)$ (vanilla gradient descent); Theorem~\ref{thm: tracking} covers only this case, while natural-gradient and momentum methods~\cite{KiBa17, StEtAl20} require separate stability analyses.
Overall, the procedure performs $T$ parameter shifts in $\lambda$ and $K$ optimization steps per shift, leading to a total of $\nupdate=TK$ optimization updates.

For any operator $A$, $\|A\|_{\rm op}$ represents its operator norm.
If $A$ depends on $\lambda$, we define $\|A\|_{\rm op} = \max_{\lambda} \|A(\lambda)\|_{\rm op}$.
For a Hermitian operator $A$, its standard deviation with respect to $\ket{\psi_t}$ is defined as
\begin{equation}\label{eq:StD}
\Std_{\psi_t}(A)=\sqrt{\bra{\psi_t} A^2\ket{\psi_t}-\big(\bra{\psi_t} A\ket{\psi_t}\big)^2}.
\end{equation}

Throughout the paper, we use the following assumptions.
\begin{description}
\item[Assumption~1 (Spectral gap)]
The instantaneous ground state of $H(\lambda)$ is nondegenerate and $\deltamin>0$.
\item[Assumption~2 (Exact representability)]
There exists a continuously differentiable parameter branch $\bth^\star(\lambda)$, with $\bth^\star(0)=\bth_0$, such that, for all $\lambda\in[0,1]$,
\begin{equation}
\ket{\psi(\bth^\star(\lambda))}=\ket{\psi_0(\lambda)}.
\end{equation}
\item[Assumption~3 (Nondegenerate geometry)]
The entire branch $\bth^\star(\lambda)$ is represented in one fixed subset of nonredundant Pauli-rotation angles, with all remaining angles held fixed, and optimization is performed in that subset.
In these coordinates,
\begin{equation}
\bg(\bth^\star(\lambda))\succeq\gamma I
\end{equation}
for some $\gamma>0$ and all $\lambda\in[0,1]$.
\end{description}
As defined above, $M$ is the total number of Pauli-rotation angles, so using $M$ in the bounds is conservative if fewer coordinates are retained.
In these coordinates, the curvature bound proved in Supplemental Material (SM)~\sm~is
\begin{equation}
\nabla^2_{\bth} E_\lambda(\bth^\star) \succeq 2\Delta(\lambda)\, \bg(\bth^\star) \succeq 2\gamma\deltamin\, I ,
\label{eq:hessian_metric_bound}
\end{equation}
which is, together with the above assumptions, the basis for proving two theorems that enable gap-conditioned, verifiable, variational ground-state tracking.

\prlsection{Adiabatic tracking via gradient descent}
Let us first give a brief, informal summary of the central result of our adiabatic optimization approach:
Gradient descent, as outlined in the preceding section, tracks the instantaneous ground state along the adiabatic path with $\nupdate$ that depends polynomially on $\deltamin^{-1}$, $\gamma^{-1}$, the number of variational parameters $M$, and the relevant operator norms.

\begin{theorem}[Adiabatic tracking]\label{thm: tracking}
Assume that Assumptions~1--3 hold and that the optimization protocol described above is used with the Pauli-rotation ansatz of \eref{eq:ansatz} containing $M$ parameters and vanilla gradient descent with \(\eta=1/L\), where \(L=4\|H\|_{\rm op}M\).
Then the following holds.

If the adiabatic discretization satisfies
\begin{equation}
\delta \lambda \leq \delta\lambda_{\mathrm{A}} = \frac{\gamma^2\,\deltamin^2} {16\,M^2\,\|H\|_{\rm op}\,\|\partial_\lambda H\|_{\rm op}}
\label{eq:dlambda_condition_explicit}
\end{equation}
and the number of optimization steps per slice satisfies
\begin{equation}
K \ge \frac{4\ln 2\; M\,\|H\|_{\rm op}}{\gamma\,\deltamin},
\label{eq:K_condition_explicit}
\end{equation}
then, at every slice \(t=1,\dots,T\), all iterates \(k=0,\ldots,K\) remain inside the local tracking region,
\begin{equation}
\big\|\bth_t^{(k)}-\bth^\star(\lambda_t)\big\|_2 \le \frac{\gamma\deltamin}{8\,M^{3/2}\|H\|_{\rm op}}.
\label{eq:tracking_region}
\end{equation}
\end{theorem}

Because the proof is stepwise, the same conclusion holds for any monotone partition \(0=\lambda_0<\cdots<\lambda_T=1\) with \(\lambda_t-\lambda_{t-1}\leq\delta\lambda_{\rm A}\) at every slice.

The local convexity and drift bounds are optimizer-independent, but the stated step count and complexity are proved only for vanilla gradient descent; other optimizers require separate local stability and contraction guarantees.

In SM~\sm, we prove \tref{thm: tracking} using the Polyak--{\L}ojasiewicz (P{\L}) condition~\cite{Ne18} and show that continued gradient descent at each slice converges to the ground state and requires only logarithmic overhead to reach a specified fidelity.
More precisely, if \(\delta E_{\rm init}\) is the initial energy error at a slice and \(0<\zeta<1\), fidelity at least \(1-\zeta\) is reached after
\begin{equation}
K_{\rm fid}\geq
\begin{cases}
\dfrac{L}{\gamma\deltamin} \ln\!\dfrac{\delta E_{\rm init}}{\zeta\deltamin}, & \delta E_{\rm init}>\zeta\deltamin,\\[0.4em]
0,&\delta E_{\rm init}\leq\zeta\deltamin.
\end{cases}
\label{eq:target_fidelity_steps}
\end{equation}

We also present a complementary imaginary-time time-dependent variational principle perspective~\cite{HaEtAl20_geometric}.

Choosing the largest admissible slice size (up to integer rounding) and the smallest admissible \(K\), the total number of updates scales as
\begin{equation}
\nupdate =\bigo\!\left( \frac{\|H\|_{\rm op}^2\, M^{3}\, \|\partial_\lambda H\|_{\rm op} }{ \gamma^{3}\, \deltamin^{3} } \right).
\label{eq:Nopt_scaling_explicit}
\end{equation}
Notably, the scaling with the gap is worse than the expected $\order{1/\deltamin^2}$ scaling observed in the adiabatic theorem~\cite{ZuEtAl25}.
For the parameter choice in Theorem~\ref{thm: tracking}, the observed $\order{\Delta_{\min}^{-3}}$ scaling admits an intuitive interpretation:
Adiabatic passage slows down as $T\propto\Delta_{\min}^{-2}$, while variational tracking requires repeated relaxation towards the instantaneous ground state whose cooling time is determined by the gap and scales as $\order{\deltamin^{-1}}$.

Compared to standard quantum annealing~\cite{KaNi98, HaEtAl20}, \tref{thm: tracking} has larger gap complexity.
However, AVQE can offer greater control over the schedule and additional options for hybrid quantum-classical optimization that are not available in real-time analog quantum annealing.
In particular, the gap does not determine the coherence time of the variational ansatz but rather the number of optimization steps, which can be beneficial in noisy devices, assuming that the required circuit depth is not too big.
Moreover, we can adjust the step size $\dl$ based on an approximation of the gap.
Also, we can offload the initial, easy part of the annealing evolution to a classical device and use the quantum device only if the classical variational ansatz is insufficiently expressive.

As a side result, the tracking region is trainable in the following precise sense.
In SM~\sm~we prove that, under the product-uniform perturbations defined there, each retained gradient component satisfies
\begin{equation}
\Var_X\!\left[\partial_{k} E_\lambda(\bth^\star+X)\right] \ge \frac{\gamma^4\deltamin^4}{192\,\|H\|_{\rm op}^2\,M^{4}}.
\label{eq:local_trainability_bound}
\end{equation}
For comparison, the same perturbation distribution gives the following lower bound on the variance of the energy values over the parameter samples (not the quantum variance of an energy measurement):
\begin{equation}
\Var_X\!\left[E_\lambda(\bth^\star+X)\right] \geq \frac{\gamma^6\deltamin^6} {184320\,\|H\|_{\rm op}^4M^8}.
\label{eq:local_energy_variance_main}
\end{equation}
These statements apply only to the tracking region and do not exclude barren plateaus elsewhere in parameter space.

\prlsection{Run-time verification via standard deviation}
\tref{thm: tracking} establishes sufficient conditions for tracking the instantaneous ground state along the adiabatic path.
However, it does not provide an \emph{a posteriori} certification procedure:
It offers no mechanism to assess whether the variational state obtained at a given step is close to an exact eigenstate, nor to verify that the final state coincides with the target ground state.

In this section, we introduce a run-time verification criterion based on the energy standard deviation, inspired by Ref.~\cite{KoEtAl19}.
This criterion localizes the state to a unique eigenvalue.
Combined with exact ground-state initialization, the tracking induction, and a gap lower bound, it certifies that the eigenvalue belongs to the ground-state branch.

\begin{theorem}[Run-time verification]\label{thm: certificate}
Consider the parameter trajectory initialized in the exact ground state at \(\lambda_0=0\), and evolved according to Theorem~\ref{thm: tracking}.
Assume that there exists a constant $\Delta_c>0$ such that $\Delta_c \leq \deltamin$.
If, for all $t=0,1,\ldots,T$,
\begin{equation}
\Std_{\psi_t}(H(\lambda_t)) \leq \frac{\Delta_c}{4}
\label{eq:variance_gap_condition}
\end{equation}
and, for every \(t<T\), the adiabatic step size satisfies
\begin{equation}
\delta \lambda < \dl_{\mathrm{V}} = \frac{\Delta_c/2-\Std_{\psi_t}(H(\lambda_t))} {\Std_{\psi_t}(H_\mathrm{f}-H_\mathrm{i})},
\label{eq:cert_cond_dl}
\end{equation}
then the state $\ket{\psi_t}$ is uniquely associated with the ground-state eigenbranch of $H(\lambda_t)$ and obeys the fidelity bound $\bigl|\langle \psi_0(\lambda_t)|\psi_t\rangle\bigr|^2\ge \frac{8}{9}.$
\end{theorem}

For a monotone adaptive partition, the step condition is imposed separately with \(\delta\lambda=\lambda_{t+1}-\lambda_t\) for every \(t<T\).

The theorem tells us that, if a gap lower bound $\Delta_c \leq \deltamin$ is available, then the standard deviation condition $\Std_{\psi_t}(H(\lambda_t))\leq\Delta_c/4$ provides a sufficient, run-time-verifiable certificate that the variational state approximates the actual ground state with fidelity at least $8/9$. Starting from the exact ground state, the half-gap step-size condition preserves the branch across slices; PŁ descent remains in the ground-state basin, and the output test certifies the optimized state.
The proof of \tref{thm: certificate} is presented in SM~\sm.
With finite shots, we replace each standard deviation by the one-sided upper confidence bound defined in Sec.~S7 of SM~\sm~and add a measured energy test.
The certificate involves no curvature or metric constants: its only spectral input is the gap bound \(\Delta_c\).
Moreover, the trainability constants entering \(\dl_{\mathrm A}\) and \(K\) need not be known at run-time--an adaptive halving schedule retains the certification guarantees unconditionally and, under Assumptions~1--3, terminates after logarithmically many restarts (see Secs.~S4 and S7 of SM~\sm)--so they enter only the complexity analysis.

Energy-based approximate representability with error \(\epsilon\) still yields the verification guarantees, provided the certificate holds.
A sufficient condition for the variational optimum to satisfy the quarter-gap test is \(\epsilon<\Delta_c^2/(16W)\), where \(W=\max_\lambda(E_{\max}(\lambda)-E_0(\lambda))\).
In contrast, tracking requires a generalized version of Assumption~3 that does not follow from the energy error alone: a smooth branch of variational optima with local curvature \(2\mu_{\mathrm{opt}}\); the tracking theorem then applies unchanged with \(\gamma\deltamin\) replaced by \(\mu_{\mathrm{opt}}\).
For \(\epsilon\leq\gamma^2\deltamin^2/(1600\,M^2W)\), metric nondegeneracy along the optimum branch implies \(\mu_{\mathrm{opt}}=\gamma\deltamin/2\)~\sm.

\prlsection{Algorithm}
The tracking and verification components combine naturally into an adiabatic variational quantum optimization protocol.
Algorithm~\ref{alg:vvqat} explicitly implements the guarantees of Theorems~1 and~2:
The optimization steps enforce adiabatic tracking, and the energy standard deviation measurements provide run-time verification (i.e., we are tracking an eigenstate with a lower bound on the fidelity) and gap-conditional certification (i.e., if the standard deviation is sufficiently small relative to the gap, then the tracked state is the ground state).

\begin{algorithm}[H]
\caption{Self-verifying AVQE}
\label{alg:vvqat}
\begin{algorithmic}[1]
\Require Initial Hamiltonian $H_\text{i}$, final Hamiltonian $H_\text{f}$, initial parameters $\bth_0$, adiabatic tracking step size $\dl_\mathrm{A}$, learning rate $\eta$, number of gradient steps $K$, estimated gap bound $\Delta_c$, safety factor $\kappa\in(0,1)$
\Ensure Final variational state $\ket{\psi(\bth)}$
\State Prepare $\ket{\psi(\bth_0)}$ as the ground state of $H(0)$
\State $\lambda \gets 0$
\State Measure $\Std_{\psi}(H(\lambda))$
\While{$\lambda < 1$}
    \State Measure $\Std_{\psi}(H_\text{f}-H_\text{i})$
    \State $\dl_{\mathrm{V}} \gets \kappa\frac{\Delta_c/2-\Std_{\psi}(H(\lambda))}{\Std_{\psi}(H_\text{f}-H_\text{i})}$
    \State $\dl \gets \min\{\dl_{\mathrm{A}},\dl_{\mathrm{V}},1-\lambda\}$
    \State $\lambda \gets \lambda + \dl$
    \State Perform $K$ gradient-descent steps on $E_\lambda(\bth)$
    \State Measure $\Std_{\psi}(H(\lambda))$
    \If{$\Std_{\psi}(H(\lambda)) > \Delta_c/4$}
        \State $\dl_{\mathrm{A}} \gets \dl_{\mathrm{A}}/2$, $K \gets 2K$, \textbf{go to line 1}
    \EndIf
\EndWhile
\State \Return $\ket{\psi(\bth)}$
\end{algorithmic}
\end{algorithm}

The safety factor $\kappa$ enforces the strict step-size inequality of \eref{eq:cert_cond_dl}.
Algorithm~\ref{alg:vvqat} is stated with exact standard deviations; in the finite-shot setting, each measured standard deviation is replaced by the one-sided upper confidence bound $\overline{\sigma}$ of Sec.~S7 of SM~\sm, the output test of line~11 is also applied at $\lambda=0$ and supplemented by a measured energy test, and line~12 also doubles the shot counts.
The algorithmic parameters $K$, $\dl_\mathrm{A}$, and $\eta$ may be chosen according to \tref{thm: tracking} or treated as hyperparameters and selected via empirical optimization; only the theorem choices---or the adaptive restart schedule of line~12 with $\eta=1/L$ (Sec.~S4 of SM~\sm)---carry the stated guarantees.

The algorithm assumes knowledge of a lower bound of the gap.
In the absence of such a lower bound, one possibility is to estimate $\Delta_c$ and run the protocol starting from that estimate, successively reducing it until the energy converges.
This heuristic does not itself certify that the estimate is a lower bound on the gap.

\prlsection{Numerical experiments}
To illustrate the protocol, we consider the one-dimensional transverse-field Ising model with longitudinal field on $n$ qubits:
\begin{align}\label{eq:ti-hamiltonian}
 H(\lambda) = \lambda J\sum_{\ell=1}^{n-1} Z_{\ell} Z_{\ell+1} + g\sum_{\ell=1}^n X_{\ell} + h\sum_{\ell=1}^n Z_{\ell},
\end{align}
for $J=1$, $g=0.5$, $h=0.25$.
The variational quantum state $\ket{\psi(\bth)}$ is defined by a Hamiltonian-informed quantum circuit of 64 layers, each containing \(n\) \(X\)-, \(n\) \(Y\)-, \(n\) \(Z\)-, and \(n-1\) nearest-neighbor \(ZZ\)-rotations, and a total of $M = 64(4n-1)$ parameters~\sm.
We investigate system sizes $n \in \{7, 11, 15\}$.
We divide the adiabatic path $\lambda\in[0,1]$ into slices and optimize each slice with gradient descent, using $K=\num{50000}$ gradient steps, a learning rate $\eta=0.1$, and momentum $\alpha=0.9$.
Due to computational limitations, we choose $K=\num{10000}$ gradient steps for $n = 15$.
The momentum update is empirical and is not covered by the vanilla gradient descent complexity proof.
The simulations use noiseless gradients evaluated by the classical simulator; shot-based parameter-shift gradients are included only in the resource analysis of Sec.~S7 of SM~\sm.
The size of each slice, $\dl$, is determined in accordance with Algorithm~\ref{alg:vvqat}, i.e. $\dl \gets \min\{\dl_{\text{A}},\dl_{\text{V}},1-\lambda\}$, and is recomputed every five adiabatic slices --- we assume that the optimization remains within the convex region between computations of $\dl$.
This recomputation schedule is an empirical shortcut and is not covered by Theorem~\ref{thm: tracking}, which requires the step-size cap at every accepted slice.

\begin{figure}
\centering
\begin{overpic}[width=\linewidth]{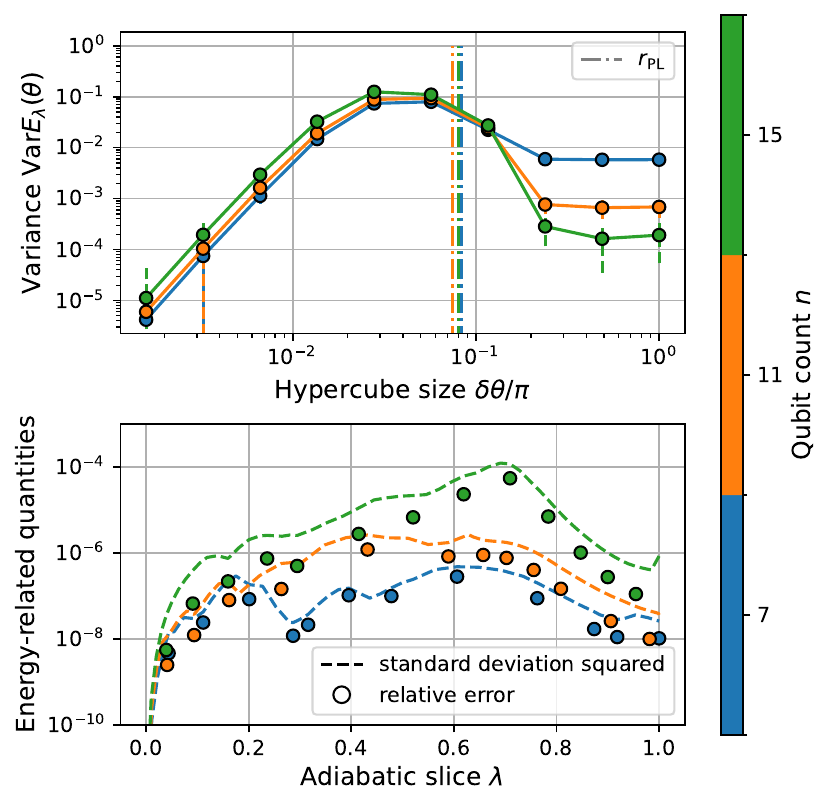}
\put(16,89){(a)}
\put(16,42){(b)}
\end{overpic}
\caption{\label{fig:2}
AVQE numerical experiments ---
(a) Energy variance with respect to the hypercube size $\delta\theta/\pi$ at the sampled slice closest to \(\lambda_\star\), where the gap is minimal.
We also plot the approximate size of the convex region, $r_{\mathrm{PL}}$, for different qubit counts.
(b) Relative energy error and squared standard deviation~\eqref{eq:StD} of the energy as a function of the adiabatic parameter $\lambda$.
}
\end{figure}

Let \(\lambda_\star\) denote the point at which the gap is minimal.
In Fig.~\ref{fig:2}(a), we report the energy variance $\text{Var}E_\lambda(\bth)$, where $\bth$ is sampled uniformly from a hypercube of size $\delta\theta$ centered at $\bth^\star$, at the sampled slice closest to \(\lambda_\star\).
The gap-dependent factors in our bounds are least favorable near this point along the adiabatic path.
We identify two regions in Fig.~\ref{fig:2}(a): for small hypercube sizes, the energy variance increases with $\delta\theta$ and does not decrease with \(n\); for large hypercube sizes, it decreases over the three sizes studied.
These finite-size data are illustrative and do not establish asymptotic scaling.
We also show the approximate size of the convex region, $r_{\text{PL}}$, for different qubit counts.
Details on how this quantity is computed can be found in Secs.~S1 and~S6 of SM~\sm.
The peak of the variance roughly aligns with $r_{\text{PL}}$.
Our results for the other sampled slices show the same qualitative behavior~\sm.

Figure~\ref{fig:2}(b) illustrates that, in these simulations, the squared standard deviation~\eqref{eq:StD} of the energy (energy variance) is closely related to the relative energy error.
It is therefore a useful run-time indicator of the accuracy of the eigenstates in these examples.

\prlsection{Shot-noise robustness}
In realistic implementations, gradients and observables are estimated from a finite number of measurement shots.
Under the shot-noise model and shot budget specified in SM~\sm, the resulting stochastic gradients remain unbiased and concentration bounds imply that, with high probability, the optimization dynamics stay within the tracking region identified in \tref{thm: tracking}.
In this regime, finite-shot noise yields a controlled optimization error floor.

The standard-deviation-based verification remains conservative under finite-shot noise when the one-sided confidence bound proved in Sec.~S7 of SM~\sm~is used.
For a variance margin of order $\Delta_c^2$, $\order{\Delta_c^{-4}}$ shots per measured Pauli term suffice, up to coefficient-norm and logarithmic confidence factors.
The ungrouped cost per slice additionally includes every Pauli term in $H$, $H^2$, $(H_\mathrm{f}-H_\mathrm{i})$, and $(H_\mathrm{f}-H_\mathrm{i})^2$, as well as all parameter-shift gradient evaluations.

\prlsection{Discussion}
Our results establish a principled paradigm for variational quantum algorithms.
The resulting algorithm admits provable correctness guarantees.
Under the stated assumptions and run-time tests, the protocol certifies both successful tracking of the target eigenstate and, conditioned on a gap bound, preparation of the actual ground state.
It therefore supplements heuristic convergence diagnostics with a rigorous run-time certificate.
The guarantees remain conditional on the scaling of \(M\), \(\gamma^{-1}\), \(\Delta_{\min}^{-1}\), Hamiltonian norms, measurement coefficients, and representability resources; no favorable system-size scaling of these quantities is proved here.

From an optimization perspective, the framework directly tackles two central challenges in variational quantum optimization.
Firstly, adiabatic warm-starting keeps the optimization in a continuously connected, trainable tracking region.
Secondly, the combination of path-following and standard-deviation-based verification provides a test against accepting spurious local minima by enforcing consistency with an underlying physical ground-state branch throughout the evolution.

More broadly, the framework provides a rigorous foundation for developing new classes of hybrid quantum-classical algorithms.
An interesting next step is to reformulate the tools developed here for adiabatic time evolution to enable us to track eigenstates other than the ground state.
In this context, there are numerous important applications, ranging from excited-state computations to the solution of linear systems of equations~\cite{SuSoOr19, CoEtAl22, CoEtAl25}.
Another interesting next step is to explore how experimental noise and gate errors affect our algorithm and whether we can develop noise-mitigation techniques tailored to our approach.
By replacing heuristic stopping rules with principled run-time verification and gap-conditional certification, our approach opens a systematic route toward reliable variational algorithms in the NISQ and early fault-tolerant era.

\prlsection{Acknowledgments}
BŽ acknowledges financial support from the Slovenian Research and Innovation Agency (ARIS) through the project J2-60034 and the UL VIP project (KTTK21) under contract no.\ SN-ZRD/22-27/510, and computational resources by SLING, a Slovenian national supercomputing network.
The authors thank Conor Mc Keever and Matthias Rosenkranz for carefully reading the manuscript and providing helpful feedback.
We used the AI tools DeepSeek~4, Kimi~3, Claude Opus~5, Claude Fable~5, Astra, and Sol to check the text for missing definitions and inconsistencies and to assist in improving the proofs, in particular the finite-shot analysis.
The authors directed all such use, verified every resulting statement and derivation, and take full responsibility for the content of the paper.

\prlsection{Data availability}
There are no publicly available research data or software supporting this manuscript.
Requests for further information or data should be sent to the authors.

\bibliography{references}

\end{document}


\title{Supplemental Material:\\
Trainable, self-verifying variational quantum eigensolver using adiabatic warm starts}

\author{Bojan Žunkovič\orcBojan}\email{bojan.zunkovic@fri.uni-lj.si}\affA\affB
\author{Marco Ballarin\orcMarco}\affC
\author{Lewis Wright\orcLewis}\affC
\author{Michael Lubasch\orcMichael}\email{michael.lubasch@quantinuum.com}\affC

\date{October 23, 2026}

\setcounter{section}{0}
\renewcommand{\thesection}{S\arabic{section}}
\renewcommand{\thesubsection}{S\arabic{section}.\arabic{subsection}}
\renewcommand{\theequation}{S\arabic{section}.\arabic{equation}}

\maketitle

We use the setup and assumptions of the main text.
First, we prove both theorems in the main text and establish trainability within the tracking region.
Then, we show how to extend our results under the assumption of $\epsilon$-exact representability.
Next, we present details on the numerical experiments.
Finally, we carry out a measurement-cost and shot-noise analysis.

\setcounter{tocdepth}{1}
\tableofcontents

\section{Proof of Theorem 1: Geometric formulation of adiabatic tracking}
\label{sec: pl proof}

In this section, we establish \texttt{Theorem 1} of the main text by a sequence of lemmas based on exact smoothness and Lipschitz bounds arising from the particular choice of the variational ansatz.
The algorithm proceeds by first shifting the adiabatic parameter $\lambda$ by a small amount $\dl$, ensuring that the previously optimized state lies in a well-behaved region of the new landscape, thereby guaranteeing linear convergence to the unique minimizer in that region.

\medskip
\noindent
\texttt{Proof roadmap:} 
\begin{enumerate}
\item \textbf{Derivative structure and smoothness.}
\lref{lem: derivatives} gives explicit formulas for all ordered energy derivatives of the assumed variational ansatz.
The subsequent Smoothness \lref{lem: smoothness} establishes global $L$--smoothness, and the Hessian Lipschitz \lref{lem: hessian lipschitz} proves that the Hessian varies in a controlled manner with Lipschitz constant $L_H$.
These lemmas provide bounds on how quickly the properties of the local optimization landscape can change.

\item \textbf{Curvature at the minimum.}
The Hessian lower bound \lref{lem: hess lower bound} shows that at $\bth^\star(\lambda)$ the Hessian is uniformly positive definite, with a lower bound determined by the spectral gap and the nondegeneracy constant of the quantum metric.
This provides a well-behaved optimization landscape at the variational minimum.

\item \textbf{Local Polyak--Łojasiewicz (PŁ) inequality.}
The Local PŁ inequality \lref{lem: PŁ inequality} combines the curvature lower bound (\lref{lem: hess lower bound}) with Hessian Lipschitz continuity (\lref{lem: hessian lipschitz}) to identify a neighborhood in which the PŁ inequality holds.

\item \textbf{Adiabatic drift and linear contraction.}
\lref{lem:drift-bound} bounds how much $\bth^\star(\lambda)$ changes with $\lambda$, while the warm-start \lref{lem: warm-start} proves contraction at fixed $\lambda$.
Together, they yield stepwise tracking.

\item \textbf{Tracking error.}
The local warm-start \lref{lem: warm-start} and the adiabatic drift bound \lref{lem:drift-bound} are then combined to derive \texttt{Theorem 1} from the main manuscript.
We distinguish two resource allocations:
\begin{itemize}
\item \texttt{Option 1:} larger step size $\dl$ and \textit{many} optimization steps for a fixed $\lambda$ -- discussed in \sref{sec:daws1}; 
\item \texttt{Option 2:} smaller step size $\dl$ and only \textit{one} optimization step for a fixed $\lambda$ -- discussed in \sref{sec:daws2}. 
\end{itemize}
In both cases, we obtain the same scaling of total complexity, with a slightly different resource allocation.
In \texttt{option 1}, we show that $\dl\propto\deltamin^2$ and the number of gradient steps $K\propto1/\deltamin$, providing the $1/\deltamin^3$ scaling.
In \texttt{option 2}, we take only one gradient update step (i.e., $K=1$) but have to use more slices, namely $\dl\propto\deltamin^3$.
Both variants also yield the same total complexity with respect to other physical parameters, but differ slightly in their numerical constants.
\texttt{Theorem 1} of the main text is for simplicity structured to be in line with \texttt{option 1}, but it could equally well be written in terms of \texttt{option 2}.
\end{enumerate}

The derivative bounds below are first proved in all $M$ Pauli-angle coordinates and remain valid after restriction.
All positivity, distance, and optimization statements use the fixed nonredundant coordinates of Assumption~3; using $M$ in the bounds is conservative when fewer coordinates are retained.

\subsection{Lemmas and linear convergence}

\begin{lemma}[Energy derivatives]
\label{lem: derivatives}
Given the variational ansatz 
\begin{equation}
    \ket{\psi(\bth)} = U_{M-1}U_{M-2}\dots U_0\ket{0}, \quad U_j = e^{-iP_j\theta_j},
\end{equation}
where $P_j$ are Hermitian Pauli strings, let $\widetilde{P}_k = (U_{M-1}\dots U_{k+1}) P_k (U_{M-1}\dots U_{k+1})^\dagger$.
For an operator $A$, let
\begin{equation}
    \mathcal{P}_j(A) = i[\widetilde{P}_j, A].
\end{equation}
For the energy $E(\bth)=\bra{\psi(\bth)}H\ket{\psi(\bth)}$, write $\partial_k=\partial/\partial\theta_k$.
For any $q\geq1$ and any indices $k_1,\ldots,k_q\in\{0,\ldots,M-1\}$, let $k_{(1)}\geq\cdots\geq k_{(q)}$ be their non-increasing rearrangement, including repeated indices.
Then
\begin{equation}
\partial_{k_1}\cdots\partial_{k_q}E(\bth)
=
\bra{\psi(\bth)}
\mathcal P_{k_{(q)}}\cdots\mathcal P_{k_{(1)}}(H)
\ket{\psi(\bth)}
\end{equation}
and
\begin{equation}
\left|
\partial_{k_1}\cdots\partial_{k_q}E(\bth)
\right|
\leq 2^q\|H\|_{\mathrm{op}}.
\label{eq:derivative_component_bound}
\end{equation}
Moreover, defining the operator norm
\begin{equation}
\|\nabla^qE(\bth)\|_{\mathrm{op}}
=
\sup_{\|v^{(1)}\|_2=\cdots=\|v^{(q)}\|_2=1}
\left|
\sum_{k_1,\ldots,k_q}
\partial_{k_1}\cdots\partial_{k_q}E(\bth)
\prod_{s=1}^q v^{(s)}_{k_s}
\right|,
\end{equation}
we have
\begin{equation}
\|\nabla^qE(\bth)\|_{\mathrm{op}}
\leq
2^q\|H\|_{\mathrm{op}}M^{q/2}.
\label{eq:derivative_tensor_bound}
\end{equation}
\end{lemma}

\begin{proof}
Direct differentiation gives
\(
\partial_k\ket{\psi}=-i\widetilde P_k\ket{\psi}
\),
and \(\partial_\ell\widetilde P_k=0\) whenever \(\ell\leq k\).
Since \(E\) is smooth, its mixed derivatives can be reordered.
After sorting the indices, differentiate successively in the order \(k_{(1)},\ldots,k_{(q)}\).
At the \(r\)-th step, every previously introduced \(\widetilde P_{k_{(s)}}\), \(s<r\), is independent of \(\theta_{k_{(r)}}\).
Thus, induction gives the stated ordered nested-commutator formula, also when indices repeat.

Since \(\|\widetilde P_j\|_{\mathrm{op}}=1\),
\[
\|\mathcal P_j(A)\|_{\mathrm{op}}
\leq2\|A\|_{\mathrm{op}}.
\]
Iteration and the bound of an expectation value by the operator norm give \eref{eq:derivative_component_bound}.
Finally, for unit vectors \(v^{(1)},\ldots,v^{(q)}\),
\[
\left|
\nabla^qE\!\left[v^{(1)},\ldots,v^{(q)}\right]
\right|
\leq
2^q\|H\|_{\mathrm{op}}
\prod_{s=1}^q\|v^{(s)}\|_1
\leq
2^q\|H\|_{\mathrm{op}}M^{q/2},
\]
which proves \eref{eq:derivative_tensor_bound}.
\end{proof}

\begin{lemma}[Smoothness]
\label{lem: smoothness}
The energy $E_\lambda(\bth)$ is $L$-smooth with
\begin{align}
L = \sup_{\bth} \| \nabla^2_\bth E(\bth) \|_{\mathrm{op}} \le 4 \|H\|_{\mathrm{op}} M ,
\end{align}
where with $\nabla_\bth E(\bth)$ we denote the vector of gradients of the energy, with elements $\big(\nabla_\bth E(\bth)\big)_k=\partial_{\theta_k}E(\bth)$.
\end{lemma}

\begin{proof}
This is \eref{eq:derivative_tensor_bound} with \(q=2\).
All statements below remain valid with \(L\) replaced by any upper bound on this supremum; Theorem~1 of the main text uses \(L=4\|H\|_{\mathrm{op}}M\).
\end{proof}

\begin{lemma}[Hessian Lipschitz continuity]
\label{lem: hessian lipschitz}
For all $\bth,\bth' \in \mathbb{R}^M$, the Hessian of the variational energy satisfies
\begin{align}
\|\nabla^2_\bth E(\bth') - \nabla^2_\bth E(\bth)\|_{\mathrm{op}} \le L_H \|\bth' - \bth\|_2 ,
\end{align}
with Lipschitz constant
\begin{align}
L_H = 8\|H\|_{\mathrm{op}} M^{3/2}.
\end{align}
\end{lemma}

\begin{proof}
Let \(\bth(t)=\bth+t(\bth'-\bth)\).
For unit vectors \(u,v\), the fundamental theorem of calculus and \eref{eq:derivative_tensor_bound} with \(q=3\) give
\[
\left|
u^\top\!\left(\nabla^2E(\bth')-\nabla^2E(\bth)\right)v
\right|
\leq
\int_0^1
\left|
\nabla^3E(\bth(t))[u,v,\bth'-\bth]
\right|dt
\leq
8\|H\|_{\mathrm{op}}M^{3/2}\|\bth'-\bth\|_2.
\]
Taking the supremum over \(u\) and \(v\) proves the claim.
\end{proof}

\begin{lemma}[Hessian lower bound] 
\label{lem: hess lower bound}
At the variational minimum $\bth^\star(\lambda)$, every vector $v$ in the fixed coordinates of Assumption~3 satisfies
\begin{equation}
v^\top\nabla^2_\bth E_\lambda(\bth^\star(\lambda))v
\geq
2\Delta(\lambda)\,v^\top\bg(\bth^\star(\lambda))v
\geq
2\gamma\deltamin\|v\|_2^2 .
\label{eq:r_pl_2}
\end{equation}
\end{lemma}

\begin{proof}
Fix \(\lambda\) and abbreviate \(H=H(\lambda)\), \(E_0=E_0(\lambda)\), and \(\bth^\star=\bth^\star(\lambda)\).
For a real vector \(v\), define the normalized curve
\[
\ket{\phi(s)}=\ket{\psi(\bth^\star+sv)},
\qquad
\ket{\phi(0)}=\ket{\psi_0}.
\]
Since \(d(\theta_\mu^\star+sv_\mu)/ds=v_\mu\), the chain rule gives
\begin{align}
\left.\frac{d^2}{ds^2}E(\bth^\star+sv)\right|_{s=0}
&=\sum_{\mu,\nu}v_\mu v_\nu
\,\partial_\mu\partial_\nu E(\bth^\star) \notag\\
&=v^\top\nabla^2_\bth E(\bth^\star)v.
\label{eq:directional_hessian}
\end{align}
Using \(E(\bth^\star+sv)=\langle\phi(s)|H|\phi(s)\rangle\), we therefore obtain
\begin{align}
v^\top\nabla^2_\bth E(\bth^\star)v
&=\left.\frac{d^2}{ds^2}
  \langle\phi(s)|H|\phi(s)\rangle\right|_{s=0} \notag\\
&=\langle\ddot\phi|H|\phi\rangle
  +2\langle\dot\phi|H|\dot\phi\rangle
  +\langle\phi|H|\ddot\phi\rangle,
\label{eq:hessian_curve_expansion}
\end{align}
where the vectors in the last line are evaluated at \(s=0\).
Twice differentiating \(\langle\phi(s)|\phi(s)\rangle=1\) gives
\[
\langle\ddot\phi|\phi\rangle
+2\langle\dot\phi|\dot\phi\rangle
+\langle\phi|\ddot\phi\rangle=0.
\]
Since \(H\ket{\phi(0)}=E_0\ket{\phi(0)}\), substituting this normalization identity into \eref{eq:hessian_curve_expansion} yields
\[
v^\top\nabla^2_\bth E(\bth^\star)v
=2\langle\dot\phi|(H-E_0I)|\dot\phi\rangle.
\]

Let \(\{\Pi_j\}_{j\geq0}\) be the spectral projectors of \(H\), with \(\Pi_0=\ketbra{\psi_0}{\psi_0}\).
The spectral gap gives
\begin{align}
H-E_0I
&=\sum_{j\geq1}(E_j-E_0)\Pi_j \notag\\
&\succeq\Delta(\lambda)\sum_{j\geq1}\Pi_j
=\Delta(\lambda)(I-\Pi_0).
\end{align}
Therefore,
\begin{align}
v^\top\nabla^2_\bth E(\bth^\star)v
&\geq2\Delta(\lambda)
\left(\|\dot\phi\|^2
-|\langle\psi_0|\dot\phi\rangle|^2\right).\label{eq:hessian_gap_tangent}
\end{align}
Finally, \(\ket{\dot\phi}=\sum_\mu v_\mu\ket{\partial_\mu\psi}\), so the definition of the quantum metric gives
\[
\|\dot\phi\|^2-|\langle\psi_0|\dot\phi\rangle|^2
=v^\top\bg(\bth^\star)v.
\]
Combining this identity with \eref{eq:hessian_gap_tangent}, Assumption~3, and \(\Delta(\lambda)\geq\deltamin\) proves
\[
v^\top\nabla^2_\bth E(\bth^\star)v
\geq2\Delta(\lambda)v^\top\bg(\bth^\star)v
\geq2\gamma\deltamin\|v\|_2^2.
\]
\end{proof}

For the remainder of this section, set \(\mu=\gamma\deltamin\).

\begin{lemma}[Local PŁ inequality]
\label{lem: PŁ inequality}
Let $\bth^\star$ denote the variational minimum at fixed $\lambda$.
For
\begin{align}\label{eq:r_pl_1}
\|\bth - \bth^\star\|_2 \le r_{\mathrm{PL}} = \frac{\mu}{L_H} \quad \mathrm{and} \quad \mu=\gamma\deltamin,
\end{align}
the PŁ inequality holds:
\begin{align}
\frac{1}{2}\|\nabla_\bth E(\bth)\|_2^2 \ge \mu \big( E(\bth) - E(\bth^\star) \big).
\end{align}
\end{lemma}

\begin{proof}
We begin by recalling from \lref{lem: hess lower bound} that the Hessian at the minimum satisfies
\begin{align}
\nabla^2_\bth E(\bth^\star) \succeq 2\mu I .
\end{align}
By \lref{lem: hessian lipschitz}, the Hessian is Lipschitz continuous with constant $L_H$, meaning
\begin{align}
\|\nabla^2_\bth E(\bth) - \nabla^2_\bth E(\bth^\star)\|_{\mathrm{op}} \le L_H \|\bth - \bth^\star\|_2 .
\end{align}
Since the Hessian is symmetric, we obtain a lower bound at $\bth$:
\begin{align}
\nabla^2_\bth E(\bth) \succeq \nabla^2_\bth E(\bth^\star) - \|\nabla^2_\bth E(\bth) - \nabla^2_\bth E(\bth^\star)\|_{\mathrm{op}} I .
\end{align}
Substituting the previous bounds yields
\begin{align}
\nabla^2_\bth E(\bth) \succeq \big( 2\mu - L_H \|\bth - \bth^\star\|_2 \big) I .
\end{align}
If $\|\bth - \bth^\star\|_2 \le \mu/L_H$, then
\begin{align}
2\mu - L_H \|\bth - \bth^\star\|_2 \ge \mu ,
\end{align}
and hence
\begin{align}
\nabla^2_\bth E(\bth) \succeq \mu I .
\end{align}
Therefore, the energy is $\mu$--strongly convex in this neighborhood.
Since the ball is convex and contains the minimizer $\bth^\star$, strong convexity implies the following bound:
\begin{align}
E(\bth) - E(\bth^\star) \le \frac{1}{2\mu} \|\nabla_\bth E(\bth)\|_2^2 .
\end{align}
Rearranging this inequality yields
\begin{align}
\frac{1}{2}\|\nabla_\bth E(\bth)\|_2^2 \ge \mu \big( E(\bth) - E(\bth^\star) \big),
\end{align}
which is exactly the PŁ inequality.
Since all steps hold whenever $\|\bth - \bth^\star\|_2 \le r_{\mathrm{PL}}$, the claim follows.
\end{proof}

\begin{lemma}[Drift bound]
\label{lem:drift-bound}
For any \(\lambda,\lambda'\in[0,1]\),
\begin{equation}
\|\bth^\star(\lambda')-\bth^\star(\lambda)\|_2
\leq D|\lambda'-\lambda|,
\qquad
D=\frac{\sqrt M\,\|\partial_\lambda H\|_{\rm op}}{\mu}.
\end{equation}
\end{lemma}

\begin{proof}
Differentiating \(\nabla_\bth E_\lambda(\bth^\star(\lambda))=0\) gives
\[
\nabla^2_\bth E_\lambda(\bth^\star)\dot{\bth}^\star
+\nabla_\bth\partial_\lambda E_\lambda(\bth^\star)=0.
\]
The Hessian lower bound and the \(q=1\) case of \eref{eq:derivative_tensor_bound}, applied to \(\partial_\lambda H\), yield
\[
\|\dot{\bth}^\star(\lambda)\|_2
\leq
\frac{\|\nabla_\bth\partial_\lambda E_\lambda(\bth^\star)\|_2}
     {2\mu}
\leq
\frac{\sqrt M\,\|\partial_\lambda H\|_{\rm op}}
     {\mu}
=D.
\]
Integration along the \(C^1\) branch proves the result.
\end{proof}

\begin{lemma}[Adiabatic warm-start convergence]
\label{lem: warm-start}
Fix \(\lambda\) and suppose \(\|\bth^{(0)}-\bth^\star(\lambda)\|_2\leq r_{\rm PL}\).
Vanilla gradient descent with step size \(1/L\) remains in this ball and satisfies
\begin{align}
\|\bth^{(k)}-\bth^\star(\lambda)\|_2
&\leq\left(1-\frac{\mu}{L}\right)^k
\|\bth^{(0)}-\bth^\star(\lambda)\|_2, \\
E_\lambda(\bth^{(k)})-E_\lambda(\bth^\star)
&\leq\left(1-\frac{\mu}{L}\right)^k\delta E_{\rm init},
\end{align}
where \(\delta E_{\rm init}=E_\lambda(\bth^{(0)})-E_\lambda(\bth^\star)\).
Consequently, if \(\delta E_{\rm init}>\epsilon_{\rm tol}>0\), energy error at most \(\epsilon_{\rm tol}\) is reached for
\[
K\geq\frac{L}{\mu}\ln\frac{\delta E_{\rm init}}{\epsilon_{\rm tol}},
\qquad
\|\bth^{(K)}-\bth^\star\|_2
\leq\sqrt{\frac{2\epsilon_{\rm tol}}{\mu}}.
\]
If \(\delta E_{\rm init}\leq\epsilon_{\rm tol}\), no additional step is required.
Choosing \(\epsilon_{\rm tol}=\zeta\deltamin\) gives
\[
\big|\langle\psi_0(\lambda)|\psi(\bth^{(K)})\rangle\big|^2
\geq1-\zeta.
\]
\end{lemma}

\begin{proof}
Inside the PŁ ball, \(E_\lambda\) is \(\mu\)-strongly convex and \(L\)-smooth.
For \(d_k=\bth^{(k)}-\bth^\star\), the fundamental theorem of calculus gives
\[
\nabla_\bth E(\bth^{(k)})=A_kd_k,\qquad
A_k=\int_0^1\nabla_\bth^2E(\bth^\star+s d_k)\,ds .
\]
The line segment lies inside the PŁ ball, so \(\mu I\preceq A_k\preceq LI\).
Therefore
\[
\|d_{k+1}\|_2
\leq\left(1-\frac{\mu}{L}\right)\|d_k\|_2.
\]
This proves invariance and distance contraction.

For the energy error \(\delta E_k=E_\lambda(\bth^{(k)})-E_\lambda(\bth^\star)\), the descent lemma at step size \(1/L\) and the PŁ inequality give
\begin{align}
\delta E_{k+1}
&\leq\delta E_k-\frac{1}{2L}
\|\nabla_\bth E_\lambda(\bth^{(k)})\|_2^2 \notag\\
&\leq\left(1-\frac{\mu}{L}\right)\delta E_k.
\end{align}
Iteration proves the stated energy contraction.
Moreover, \((1-\mu/L)^K\leq e^{-K\mu/L}\), so the stated lower bound on \(K\) implies \(\delta E_K\leq\epsilon_{\rm tol}\).
Strong convexity then yields
\[
\frac{\mu}{2}\|\bth^{(K)}-\bth^\star\|_2^2
\leq\delta E_K\leq\epsilon_{\rm tol},
\]
which proves the parameter-distance bound.

It remains to convert the energy error into fidelity.
Exact representability gives \(E_\lambda(\bth^\star)=E_0(\lambda)\).
Expanding \(\ket{\psi(\bth^{(K)})}\) in the eigenbasis of \(H(\lambda)\), with weights \(p_j\), gives
\begin{align}
\delta E_K
&=\sum_{j\geq1}p_j\big(E_j(\lambda)-E_0(\lambda)\big) \notag\\
&\geq\Delta(\lambda)\sum_{j\geq1}p_j \notag\\
&=\Delta(\lambda)
\left(1-\big|\langle\psi_0(\lambda)|
\psi(\bth^{(K)})\rangle\big|^2\right).
\end{align}
Since \(\Delta(\lambda)\geq\deltamin\), the choice \(\epsilon_{\rm tol}=\zeta\deltamin\) and \(\delta E_K\leq\epsilon_{\rm tol}\) imply the stated fidelity bound.
\end{proof}

\subsection{Ground state tracking --- option 1}
\label{sec:daws1}

We discretize $\lambda$ so that consecutive minimizers lie within half of the PŁ radius.
By \lref{lem:drift-bound}, this is ensured if
\begin{equation}
\delta\lambda
\leq\delta\lambda_{\rm A,1}
:=
\frac{\gamma^2\deltamin^2}
{2L_H\sqrt M\,\|\partial_\lambda H\|_{\rm op}}.
\label{eq:delta-lambda-bound}
\end{equation}
The right-hand side is interpreted as $+\infty$ when $\|\partial_\lambda H\|_{\rm op}=0$.

Assume that at step $t$ the iterate satisfies
\[
\|\bth_t-\bth^\star(\lambda_t)\|_2\leq\frac{r_{\rm PL}}2,
\]
then
\[
\begin{aligned}
\|\bth_t-\bth^\star(\lambda_{t+1})\|_2
&\leq
\|\bth_t-\bth^\star(\lambda_t)\|_2
+\|\bth^\star(\lambda_t)-\bth^\star(\lambda_{t+1})\|_2\\
&\leq r_{\rm PL}.
\end{aligned}
\]
Hence, the warm start always lies inside the PŁ region of the next energy landscape.

Inside the PŁ region, by \lref{lem: warm-start}, gradient descent with step size $\eta=1/L$ reduces the parameter error by a factor of $2$ whenever
\begin{equation}
K\geq\frac{L}{\gamma\deltamin}\ln2.
\label{eq:gd-steps}
\end{equation}
Because \(\bth_0=\bth^\star(0)\), the output therefore returns to the inner half of the PŁ region at every slice.
The same induction applies to a nonuniform monotone partition by taking \(\delta\lambda=\lambda_{t+1}-\lambda_t\) at each step.

Substituting \(L=4\|H\|_{\rm op}M\) and \(L_H=8\|H\|_{\rm op}M^{3/2}\) gives the constants in Theorem~1,
\[
\delta\lambda_{\rm A,1}
=
\frac{\gamma^2\deltamin^2}
{16M^2\|H\|_{\rm op}\|\partial_\lambda H\|_{\rm op}},
\qquad
K\geq
\frac{4\ln2\,M\|H\|_{\rm op}}{\gamma\deltamin}.
\]
Choosing \(T=\lceil1/\delta\lambda_{\rm A,1}\rceil\) gives
\[
\nupdate
=\bigo\!\left(
\frac{\|H\|_{\rm op}^2M^3\|\partial_\lambda H\|_{\rm op}}
{\gamma^3\deltamin^3}
\right),
\]
with leading prefactor \(64\ln2\) before integer rounding.
This is the protocol used in \texttt{Theorem 1} of the main manuscript.
An alternative version is presented in the following subsection.

\subsection{Ground state tracking --- option 2}
\label{sec:daws2}

In this section, we refine the warm-start analysis of \sref{sec:daws1} by explicitly tracking the \emph{transient optimization error} across adiabatic steps.
Rather than requiring near-exact re-optimization at each step, we show that finite optimization effort per step suffices: the tracking error obeys a contractive recursion with an $\bigo(\dl)$ forcing term, yielding a uniform steady-state bound.
In the end, we achieve the same level of optimization complexity with a slightly different analysis.

Let \(\bth_t\) denote the output after \(K\) gradient steps at \(\lambda_t\), and set
\[
\mu=\gamma\deltamin,\qquad
r_{\rm PL}=\frac{\mu}{L_H},\qquad
D=\frac{\sqrt M\,\|\partial_\lambda H\|_{\rm op}}{\mu},
\qquad
q=1-\frac{\mu}{L},\qquad
a=q^K.
\]
Here \(0\leq q<1\).
Whenever the warm start lies in the PŁ ball, \lref{lem:drift-bound} and \lref{lem: warm-start} give
\begin{equation}
\|e_{t+1}\|_2
\leq
a\bigl(\|e_t\|_2+D\delta\lambda\bigr),
\qquad
e_t=\bth_t-\bth^\star(\lambda_t).
\label{eq: forced contraction}
\end{equation}

Provided that the warm starts up to slice \(t\) lie in the PŁ region, iterating \eref{eq: forced contraction} gives
\begin{equation}
\|e_t\|_2
\leq
a^t\|e_0\|_2
+\frac{a(1-a^t)}{1-a}D\delta\lambda.
\label{eq: uniform error bound}
\end{equation}

The explicit bound above is conditional: \eref{eq: forced contraction} is available at the next slice only if the pre-optimization warm-start error \(\|e_t\|_2+D\delta\lambda\) does not exceed \(r_{\rm PL}\).
Define
\[
R=\frac{a}{1-a}D\delta\lambda.
\]
It is sufficient to impose
\begin{equation}
D\delta\lambda
\leq
(1-a)r_{\rm PL}.
\label{eq:option2_step_condition}
\end{equation}
because then
\[
R+D\delta\lambda
=\frac{D\delta\lambda}{1-a}
\leq r_{\rm PL}.
\]
Indeed, \(e_0=0\leq R\).
If \(\|e_t\|_2\leq R\), the next warm start is in the PŁ region and \eref{eq: forced contraction} gives
\[
\|e_{t+1}\|_2
\leq a(R+D\delta\lambda)=R.
\]
Induction, therefore, proves both the validity of the recursion and uniform tracking.
By contrast, the condition \(D\delta\lambda<(1-a)r_{\rm PL}/a\) guarantees only \(R<r_{\rm PL}\); it does not guarantee \(R+D\delta\lambda\leq r_{\rm PL}\).

For fixed \(K\geq1\), define the admissible cap
\[
\delta\lambda_{\max}
=
\frac{(1-q^K)r_{\rm PL}}{D}.
\]
If $D=0$, there is no minimizer drift, and no cap is needed; below, assume $D>0$.
Choosing $T=\lceil1/\delta\lambda_{\max}\rceil$ and the uniform step $1/T$ gives
\[
\nupdate
=
K\left\lceil\frac1{\delta\lambda_{\max}}\right\rceil
\leq
K+\frac{KD}{(1-q^K)r_{\rm PL}}.
\]
Since \(1-q^K\geq1-q=\mu/L\), this has the same \(O(\deltamin^{-3})\) scaling as Option~1 for fixed \(K\).

For one gradient step per slice,
\begin{equation}
\delta\lambda
\leq\delta\lambda_{\rm A,2}
:=
\frac{\mu r_{\rm PL}}{LD}
=
\frac{\gamma^3\deltamin^3}
{32M^3\|H\|_{\rm op}^2
\|\partial_\lambda H\|_{\rm op}}.
\label{eq:option2_explicit_step}
\end{equation}
Choosing \(T=\lceil1/\delta\lambda_{\rm A,2}\rceil\) gives
\[
\nupdate
=
\left\lceil\frac1{\delta\lambda_{\rm A,2}}\right\rceil
\leq
1+
\frac{32M^3\|H\|_{\rm op}^2
\|\partial_\lambda H\|_{\rm op}}
{\gamma^3\deltamin^3}.
\]

\section{Complementary continuous-time adiabatic formulation and scaling}

This section gives a complementary continuous-time time-dependent variational principle (TDVP) bound.
It is not used in the discrete update-count proof of \texttt{Theorem 1}.

We use the fixed coordinates of Assumption~3 and quantify below a neighborhood in which the quantum metric remains invertible.

In conventional imaginary time \(t\), the projective McLachlan TDVP equation in our metric convention is~\cite{HaEtAl20_geometric}
\[
\bg(\bth)\frac{d\bth}{dt}=-\frac12\nabla_\bth E.
\]
Setting \(\tau=t/2\) and rescaling the total duration accordingly gives, for the schedule \(\lambda(\tau)=\tau/T\),
\begin{align}
\frac{d\bth}{d\tau}
=-\bg(\bth)^{-1}\nabla_\bth E_{\lambda(\tau)}(\bth).
\label{eq:tdvp-flow}
\end{align}
Throughout this section, a dot denotes a derivative with respect to \(\tau\):
\[
\dot{\bth}=\frac{d\bth}{d\tau},\qquad
\dot{\delta\bth}=\frac{d\delta\bth}{d\tau},\qquad
\dot\lambda=\frac{d\lambda}{d\tau}=\frac1T.
\]

Let \(\delta\bth=\bth-\bth^\star(\lambda(\tau))\), \(\bg_\star=\bg(\bth^\star(\lambda))\), and \(\mathcal H=\nabla^2_\bth E_\lambda(\bth^\star(\lambda))\).
Taylor expansion of Eq.~\eqref{eq:tdvp-flow} gives
\begin{equation}
\dot{\delta\bth}
=-\bg_\star^{-1}\mathcal H\delta\bth
-\frac1T\partial_\lambda\bth^\star
+\mathcal R(\delta\bth,\lambda),
\label{eq: tdvp reminder}
\end{equation}
with the nonlinear remainder
\[
\mathcal R(\delta\bth,\lambda)
:=-\bg(\bth^\star+\delta\bth)^{-1}
\nabla_\bth E_\lambda(\bth^\star+\delta\bth)
+\bg_\star^{-1}\mathcal H\delta\bth.
\]
Since \(\nabla_\bth E_\lambda(\bth^\star)=0\), smoothness gives \(\mathcal R=O(\|\delta\bth\|_2^2)\) uniformly along the compact branch.

We measure the tracking error in the instantaneous quantum metric by
\[
u^2=\delta\bth^\top\bg_\star\delta\bth.
\]
Writing \(\|x\|_{\bg_\star}=(x^\top\bg_\star x)^{1/2}\), let \(L_g\) be a uniform Lipschitz constant of the quantum metric:
\[
\|\bg(\bth)-\bg(\bth')\|_{\rm op}
\leq L_g\|\bth-\bth'\|_2.
\]
For the Pauli-rotation ansatz, \(\|\partial_j\psi\|,\|\partial_k\partial_j\psi\|\leq1\).
Write
\[
g_{ij}=\operatorname{Re}(A_{ij}-a_i b_j),\quad
A_{ij}=\langle\partial_i\psi|\partial_j\psi\rangle,\quad
a_i=\langle\partial_i\psi|\psi\rangle,\quad
b_j=\langle\psi|\partial_j\psi\rangle.
\]
Differentiating \(A_{ij}\) gives two inner products, each bounded by one, and therefore \(|\partial_k A_{ij}|\leq2\).
Likewise, \(|\partial_k a_i|,|\partial_k b_j|\leq2\) and \(|a_i|,|b_j|\leq1\), so the product rule gives
\[
|\partial_k(a_i b_j)|
\leq|\partial_k a_i|\,|b_j|+|a_i|\,|\partial_k b_j|
\leq4.
\]
Thus \(|\partial_k g_{ij}|\leq2+4=6\).
Converting this entrywise bound to the operator norm shows that we may take \(L_g=6M^{3/2}\).

We now choose the tracking-tube radius
\begin{equation}
\rho
:=\sqrt\gamma\min\left\{r_{\rm PL},\frac{\gamma}{2L_g}\right\}
=\gamma^{3/2}\min\left\{
\frac{\deltamin}{L_H},\frac{1}{12M^{3/2}}
\right\}
=\frac{\gamma^{3/2}}{M^{3/2}}
\min\left\{\frac{\deltamin}{8\|H\|_{\rm op}},\frac1{12}\right\}.
\label{eq:tdvp_tube_radius}
\end{equation}
Indeed, Assumption~3 gives \(\|\delta\bth\|_2\leq u/\sqrt\gamma\).
Hence \(u\leq\rho\) implies both \(\|\delta\bth\|_2\leq r_{\rm PL}\) and
\begin{align*}
\bg(\bth^\star+\delta\bth)
&\succeq\bg_\star-L_g\|\delta\bth\|_2 I\\
&\succeq\frac{\gamma}{2}I.
\end{align*}
Thus, the tube is contained in the PŁ neighborhood, and the TDVP metric is uniformly invertible there.

Applying \lref{lem: hess lower bound} with \(v=\delta\bth\), and using \(u^2=\delta\bth^\top\bg_\star\delta\bth\) and \(\Delta(\lambda)\geq\deltamin\), gives
\[
\delta\bth^\top\mathcal H\delta\bth
\geq2\Delta(\lambda)\,
\delta\bth^\top\bg_\star\delta\bth
=2\Delta(\lambda)u^2
\geq2\deltamin u^2.
\]
For \(0<u\leq\rho\), differentiating \(u^2=\delta\bth^\top\bg_\star\delta\bth\) and using the symmetry of \(\bg_\star\) gives
\begin{align*}
2u\dot u
&=\dot{\delta\bth}^\top\bg_\star\delta\bth
+\delta\bth^\top\bg_\star\dot{\delta\bth}
+\delta\bth^\top\frac{d\bg_\star}{d\tau}\delta\bth\\
&=2\delta\bth^\top\bg_\star\dot{\delta\bth}
+\delta\bth^\top\frac{d\bg_\star}{d\tau}\delta\bth,\\
u\dot u
&=\delta\bth^\top\bg_\star\dot{\delta\bth}
+\frac{1}{2T}\delta\bth^\top
\frac{d\bg_\star}{d\lambda}\delta\bth,
\end{align*}
where we used \(d\bg_\star/d\tau=(d\bg_\star/d\lambda)\dot\lambda\) and \(\dot\lambda=1/T\).
Substituting Eq.~\eqref{eq: tdvp reminder} then gives
\begin{align*}
u\dot u
&=-\delta\bth^\top\mathcal H\delta\bth
-\frac1T\delta\bth^\top\bg_\star
\partial_\lambda\bth^\star
+\delta\bth^\top\bg_\star\mathcal R(\delta\bth,\lambda)
+\frac{1}{2T}\delta\bth^\top
\frac{d\bg_\star}{d\lambda}\delta\bth.
\end{align*}
Here, the first term follows from \(\bg_\star\bg_\star^{-1}=I\).
The last three terms satisfy
\begin{align*}
-\delta\bth^\top\bg_\star\partial_\lambda\bth^\star
&\leq\left|\delta\bth^\top\bg_\star
\partial_\lambda\bth^\star\right|
\\&\leq
\|\delta\bth\|_{\bg_\star}
\|\partial_\lambda\bth^\star\|_{\bg_\star}
\leq Bu,\\
\delta\bth^\top\bg_\star\mathcal R(\delta\bth,\lambda)
&\leq\left|\delta\bth^\top\bg_\star
\mathcal R(\delta\bth,\lambda)\right|
\\&\leq
\|\delta\bth\|_{\bg_\star}
\|\mathcal R(\delta\bth,\lambda)\|_{\bg_\star}
\leq C_{\rm nl}u^3,\\
\left|\delta\bth^\top
\frac{d\bg_\star}{d\lambda}\delta\bth\right|
&=\left|\left(\bg_\star^{1/2}\delta\bth\right)^\top
\left(\bg_\star^{-1/2}\frac{d\bg_\star}{d\lambda}
\bg_\star^{-1/2}\right)
\left(\bg_\star^{1/2}\delta\bth\right)\right|\\
&\leq
\left\|\bg_\star^{-1/2}\frac{d\bg_\star}{d\lambda}
\bg_\star^{-1/2}\right\|_{\rm op}u^2
\leq\beta u^2,
\end{align*}
where the uniform constants are
\begin{align*}
B&:=\sup_{\lambda\in[0,1]}
\|\partial_\lambda\bth^\star\|_{\bg_\star},\\
C_{\rm nl}
&:=\sup_{\substack{\lambda\in[0,1],\,\delta\bth\\
0<\delta\bth^\top\bg_\star\delta\bth\leq\rho^2}}
\frac{\|\mathcal R(\delta\bth,\lambda)\|_{\bg_\star}}
{\delta\bth^\top\bg_\star\delta\bth},\\
\beta
&:=\sup_{\lambda\in[0,1]}
\left\|
\bg_\star^{-1/2}\frac{d\bg_\star}{d\lambda}\bg_\star^{-1/2}
\right\|_{\rm op}.
\end{align*}
They are finite by the smoothness and compactness of the branch.
Here \(B\) is the maximum speed of the variational minimum, \(C_{\rm nl}\) controls the nonlinear TDVP terms, and \(\beta\) controls the variation of the metric.
Combining these estimates with the Hessian bound above yields
\[
u\dot u
\leq-2\deltamin u^2+\frac{B}{T}u
+C_{\rm nl}u^3+\frac{\beta}{2T}u^2.
\]
Dividing by \(u\) gives
\begin{equation}
\dot u
\leq
-\left(2\deltamin-\frac{\beta}{2T}\right)u
+\frac{B}{T}
+C_{\rm nl}u^2 .
\label{eq:tdvp_barrier_inequality}
\end{equation}
At a point where \(u=0\), Eq.~\eqref{eq: tdvp reminder} directly gives the same bound for the right derivative, \(\dot u\leq B/T\).

We now construct an upper bound that remains inside the tube.
First require \(T\geq\beta/(2\deltamin)\), so that Eq.~\eqref{eq:tdvp_barrier_inequality} becomes
\[
\dot u\leq-\deltamin u+\frac{B}{T}+C_{\rm nl}u^2.
\]
If \(u\leq A\) and \(C_{\rm nl}A\leq\deltamin/2\), the nonlinear term can be absorbed into half of the linear contraction, giving
\[
\dot u\leq-\frac{\deltamin}{2}u+\frac{B}{T}.
\]
The right-hand side vanishes at
\begin{equation}
A:=\frac{2B}{T\deltamin},
\label{eq:tdvp_barrier}
\end{equation}
which motivates \(A\) as the proposed upper bound.
The requirements \(A\leq\rho\) and \(C_{\rm nl}A\leq\deltamin/2\) are satisfied when \(T\geq2B/(\rho\deltamin)\) and \(T\geq4C_{\rm nl}B/\deltamin^2\), respectively.
Thus we take \(u(0)=0\) and impose the combined condition
\begin{equation}
T\geq
\max\left\{
\frac{\beta}{2\deltamin},
\frac{4C_{\rm nl}B}{\deltamin^2},
\frac{2B}{\rho\deltamin}
\right\}.
\label{eq:tdvp_T_condition}
\end{equation}
It remains to prove that the proposed bound \(u\leq A\) is indeed preserved.
Suppose that \(u\) reaches \(A\) during \(0\leq\tau\leq T\), and let \(\tau_{\rm e}>0\) be the first time at which this happens.
On \([0,\tau_{\rm e}]\), integration of the last inequality gives
\begin{equation}
u(\tau)
\leq
\frac{2B}{T\deltamin}
\left(1-e^{-\deltamin\tau/2}\right),
\qquad 0\leq\tau\leq\tau_{\rm e} .
\label{eq:tdvp_continuous_bound}
\end{equation}
At \(\tau=\tau_{\rm e}\), the right-hand side is strictly smaller than \(A\), contradicting \(u(\tau_{\rm e})=A\).
Therefore, no such exit time exists.
Applying the same integration on \([0,T]\) proves the bound in Eq.~\eqref{eq:tdvp_continuous_bound} for \(0\leq\tau\leq T\), and \(u(\tau)<A\leq\rho\).
Thus, the trajectory remains inside the tube in which the local bounds were derived.

For the Pauli-rotation ansatz, \(\|\bg\|_{\rm op}\leq M\); hence \lref{lem:drift-bound} gives \(B\leq\sqrt M D=O(\deltamin^{-1})\).
Once condition~\eqref{eq:tdvp_T_condition} holds, the error bound is therefore \(u=O(T^{-1}\deltamin^{-2})\).
The nonlinear-stability condition may require \(T\) of order \(\deltamin^{-3}\), since \(B=O(\deltamin^{-1})\).

The continuous TDVP result bounds a physical-time tracking error subject to condition~\eqref{eq:tdvp_T_condition}; it does not count optimizer updates.
The two discrete P\L{} options both give \(\nupdate=O(\deltamin^{-3})\):
Option~1 uses \(\delta\lambda=O(\deltamin^2)\) and \(K=O(\deltamin^{-1})\), whereas Option~2 uses \(\delta\lambda=O(\deltamin^3)\) and \(K=1\).

\section{Trainability within the tracking region}

This statement applies only to the basin used by the tracking proof; it is not a global barren-plateau claim.

\begin{theorem}[Non-vanishing gradient variance in the tracking region]
Fix \(\lambda\), let \(d\) be the number of parameters retained in Assumption~3, and let \(r_{\rm PL}={\gamma\deltamin}/{L_H}\) be the radius of \eref{eq:r_pl_1}.
Let the components of \(X\in\mathbb R^d\) be independent with
\[
X_j\sim\operatorname{Unif}\!\left[
-\frac{r_{\rm PL}}{\sqrt d},\frac{r_{\rm PL}}{\sqrt d}
\right].
\]
Then the support of \(\bth^\star+X\) lies inside the tracking region and, for every \(k=1,\ldots,d\),
\begin{equation}
\operatorname{Var}_X\!\left[
\partial_kE_\lambda(\bth^\star+X)\right]
\geq
\frac{\gamma^4\deltamin^4}
{192\,d\|H\|_{\rm op}^2M^3}.
\label{eq:local_gradient_variance}
\end{equation}
\end{theorem}

\begin{proof}
Since \(\|X\|_2\leq r_{\rm PL}\), its support lies in the PŁ ball.
There \(\nabla^2E_\lambda\succeq\mu I\), with \(\mu=\gamma\deltamin\).

Write
\[
X=X_{-k}+X_ke_k,
\qquad
X_{-k}:=\sum_{j\ne k}X_je_j.
\]
Fix any value \(x_{-k}\) in the support of \(X_{-k}\), and define
\[
h(x)=\partial_kE_\lambda(\bth^\star+x_{-k}+xe_k).
\]
Throughout the support, \(h'(x)=\partial_k^2E_\lambda(\bth^\star+x_{-k}+xe_k)\geq\mu\).
The mean-value theorem therefore gives, for any \(x,y\) in the support,
\[
|h(x)-h(y)|\geq\mu|x-y|.
\]

Let \(U,U'\) be independent copies of \(X_k\).
For any random variable \(Y\) and an independent copy \(Y'\),
\[
\frac12\mathbb E_{Y,Y'}[(Y-Y')^2]
=\mathbb E_Y[Y^2]-\bigl(\mathbb E_Y[Y]\bigr)^2
=\operatorname{Var}(Y).
\]
Here \(\mathbb E_{Y,Y'}\) denotes expectation over the joint distribution of the independent pair \((Y,Y')\).
Applying this identity first to \(Y=h(U)\), then using the preceding mean-value bound, and finally applying it to \(Y=U\), gives
\begin{align*}
\operatorname{Var}(h(U))
&=\frac12\mathbb E_{U,U'}[(h(U)-h(U'))^2]\\
&\geq\frac{\mu^2}{2}\mathbb E_{U,U'}[(U-U')^2]
=\mu^2\operatorname{Var}(U)\\
&=\frac{\mu^2r_{\rm PL}^2}{3d},
\end{align*}
where the last equality uses \(\operatorname{Var}(U)=r_{\rm PL}^2/(3d)\) for the stated uniform distribution.
This bound holds for every fixed \(x_{-k}\).
Hence, writing \(G=\partial_kE_\lambda(\bth^\star+X)\), the law of total variance decomposes the variance of \(G\) into the average variance over \(X_k\) at
fixed \(X_{-k}\) and the variance of the corresponding conditional mean:
\begin{align*}
\operatorname{Var}(G)
&=\mathbb E\!\left[\operatorname{Var}(G\mid X_{-k})\right]
+\operatorname{Var}\!\left(\mathbb E[G\mid X_{-k}]\right)\\
&\geq\frac{\mu^2r_{\rm PL}^2}{3d}
=\frac{\gamma^4\deltamin^4}
{192\,d\|H\|_{\rm op}^2M^3}.
\end{align*}
The second term in the equality is nonnegative, while the first is bounded below by the fixed-\(x_{-k}\) estimate above.
This proves \eref{eq:local_gradient_variance}.
\end{proof}

Since \(d\leq M\), \eref{eq:local_gradient_variance} implies the weaker bound \(\operatorname{Var}_X[\partial_kE_\lambda(\bth^\star+X)]\geq \gamma^4\deltamin^4/(192\,\|H\|_{\rm op}^2M^4)\) stated in the main text.

\begin{theorem}[Non-vanishing energy variance in the tracking region]
Fix \(\lambda\), let \(d\) be the number of parameters retained in Assumption~3, set \(\mu=\gamma\deltamin\) and \(r_{\rm PL}=\gamma\deltamin/L_H\), and let the components of \(X\in\mathbb R^d\) be independent with
\[
X_j\sim\operatorname{Unif}\!\left[
-\frac{r_{\rm PL}}{\sqrt d},\frac{r_{\rm PL}}{\sqrt d}
\right].
\]
Then the variance over the parameter perturbation \(X\) satisfies
\begin{equation}
\operatorname{Var}_X\!\left[E_\lambda(\bth^\star+X)\right]
\geq
\frac{\gamma^6\deltamin^6}
{184320\,d^2\|H\|_{\rm op}^4M^6}.
\label{eq:local_energy_variance}
\end{equation}
\end{theorem}

\begin{proof}
Set \(a=r_{\rm PL}/\sqrt d\).
As in the preceding theorem, the support of \(X\) lies in the PŁ ball, where \(\nabla^2E_\lambda\succeq\mu I\).
Fix one coordinate \(k\) and any value \(x_{-k}\) in the support of \(X_{-k}=\sum_{j\ne k}X_je_j\), and define
\[
f(x)=E_\lambda(\bth^\star+x_{-k}+xe_k),
\qquad -a\leq x\leq a.
\]
Then \(f''(x)=\partial_k^2E_\lambda\geq\mu\).
Introduce the even part of \(f\) and the comparison parabola
\[
q(x)=\frac{f(x)+f(-x)}2,
\qquad p(x)=q(0)+\frac\mu2x^2.
\]
Here \(q\) is the even part of \(f\), while \(f-q\) is odd.
Moreover, \(q'(0)=0\) and
\[
q''(x)=\frac{f''(x)+f''(-x)}2\geq\mu.
\]
It follows that \(q'(t)\geq\mu t\) for \(t\geq0\).
Thus, for \(0\leq y\leq x\),
\[
q(x)-q(y)
=\int_y^xq'(t)\,dt
\geq\mu\int_y^xt\,dt
=p(x)-p(y).
\]
By evenness, this implies
\[
|q(x)-q(y)|\geq|p(x)-p(y)|
\qquad\text{for all }x,y\in[-a,a].
\]

Let \(U\sim\operatorname{Unif}[-a,a]\).
Since \(q(U)\) is even in \(U\) and \(f(U)-q(U)\) is odd, the two terms are uncorrelated.
If \(U'\) is an independent copy of \(U\), the independent-copy variance identity therefore gives
\begin{align*}
\operatorname{Var}(f(U))
&\geq\operatorname{Var}(q(U))\\
&=\frac12\mathbb E_{U,U'}[(q(U)-q(U'))^2]\\
&\geq\frac12\mathbb E_{U,U'}[(p(U)-p(U'))^2]
=\operatorname{Var}(p(U)).
\end{align*}
The remaining calculation is the variance of the parabola:
\[
\operatorname{Var}(p(U))
=\frac{\mu^2}{4}
\left(\frac{a^4}{5}-\frac{a^4}{9}\right)
=\frac{\mu^2a^4}{45}
=\frac{\mu^2r_{\rm PL}^4}{45d^2}.
\]
This bound holds for every fixed \(x_{-k}\).
The law of total variance over \(X_k\) and \(X_{-k}\) therefore gives the same lower bound for \(\operatorname{Var}_X[E_\lambda(\bth^\star+X)]\).
Finally, substituting \(\mu=\gamma\deltamin\), \(r_{\rm PL}=\gamma\deltamin/L_H\), and \(L_H=8\|H\|_{\rm op}M^{3/2}\) proves \eref{eq:local_energy_variance}.
\end{proof}

Using \(d\leq M\), the theorem also implies
\[
\operatorname{Var}_X\!\left[E_\lambda(\bth^\star+X)\right]
\geq
\frac{\gamma^6\deltamin^6}
{184320\,\|H\|_{\rm op}^4M^8}.
\]

\section{Proof of Theorem 2: Standard-deviation-based verification}
\label{sec:verification}

In this section, we introduce a \emph{verification layer} which, combined with exact ground-state initialization and the tracking induction, certifies at each adiabatic step that the variational state lies on the ground-state spectral branch.
The standard-deviation diagnostic itself is independent of the optimization method.
It does not drive the optimization but serves to identify a unique nearby eigenvalue and, once the branch is identified, to quantify the overlap with the corresponding eigenstate.
The identification of this branch as the ground branch is supplied by the warm-start induction in items~3--5 of the proof roadmap below.
Parameter distances and the PŁ radius appearing in this section are evaluated in the fixed coordinates of Assumption~3 of the main text, as in Sec.~\ref{sec: pl proof}.

\subsection{Setup, assumptions, and proof roadmap}

To make the presentation easier, we extend our setup and assumptions.
We define the spectral decomposition
\[
H(\lambda) = \sum_{j\ge 0} E_j(\lambda)\, \Pi_j(\lambda),
\qquad
E_0(\lambda) < E_1(\lambda) \le E_2(\lambda) \le \cdots ,
\]
where $\Pi_j(\lambda)$ denotes a projector to the $j-$th eigenstate.
For any normalized state $\ket{\psi}$ and Hermitian operator $A$, we define
\begin{align}
\Std_\psi(A)
&=  \sqrt{\bra{\psi} A^2 \ket{\psi} - \bra{\psi} A\ket{\psi}^2}, \\
E_\psi(\lambda)
&= \bra{\psi} H(\lambda) \ket{\psi}, \quad \mbox{and}\quad  \Std_\psi(\lambda)=\Std_\psi(H(\lambda)).
\end{align}
In addition, we introduce a new assumption.
\begin{assumption}[Uniform gap \emph{to the tracked branch}]
\label{ass:global_gap_ver}
Let $j_\star$ denote the eigen-branch.
Assume there exists $\deltaminj>0$ such that, for all $\lambda\in[0,1]$,
\begin{equation}
\min_{j\neq j_\star} \big|E_j(\lambda)-E_{j_\star}(\lambda)\big|
\ge
\deltaminj.
\label{eq:tracked_gap}
\end{equation}
\end{assumption}
For the main theorem, we use $j_\star=0$ (which reduces this assumption to Assumption~1 of the main text), but we state some lemmas more generally.
Every ground-branch bound below remains valid with the known lower bound \(\Delta_c\leq\deltamin\) in place of \(\deltamin\), as used in the main text.

\medskip
\noindent
\texttt{Proof roadmap:}
The goal of this section is to turn \emph{measured energy standard deviation} into a \emph{run-time verification} that the algorithm remains on the intended eigen-branch (in particular, the ground-state branch) throughout the adiabatic evolution, and to quantify the resulting fidelity.
The proof of \texttt{Theorem~2} in the main text is obtained in several steps:

\begin{enumerate}
\item \textbf{Eigen-branch identification from standard deviation.}
First, we show that, for any state $\ket{\psi}$ at a fixed $\lambda$, the mean energy $E_\psi(\lambda)$ must lie within one standard deviation of \emph{some} eigenvalue $E_j(\lambda)$.
Moreover, if this index is the tracked branch and the measured standard deviation is smaller than half the uniform branch gap, then this index is \emph{unique}.
This is \lref{lem:variance_branch}.

\item \textbf{From identification to fidelity.}
Once the relevant branch $j_\star$ is uniquely identified, we convert the same standard deviation value into a quantitative lower bound on the overlap $\bra{\psi}\Pi_{j_\star}(\lambda)\ket{\psi}$.
This is Lemma~\ref{lem:var_to_fid_branch} and its specialization to the ground state in \cref{cor:ground_fidelity}.
In our final theorem, we use $j_\star=0$.

\item \textbf{Warm-start verification for linear schedules.}
We then analyze the canonical warm-start state $\ket{0(\lambda)}$ when the schedule advances to $\lambda^+=\lambda+\delta\lambda$.
For a linear interpolation $H(\lambda)$, the energy variance of the warm start with respect to the \emph{new} Hamiltonian is \emph{exactly quadratic} in $\delta\lambda$, yielding an explicit step-size condition that guarantees $\Std_{0(\lambda)}(H(\lambda^+))<\deltamin\,/2$.
Since this bound holds throughout the step, continuity preserves the initially identified ground branch.
This is Lemma~\ref{lem:warm_start_exact} and the subsequent step-size lemma.

\item \textbf{Warm-start verification for general states.}
To decouple verification from the assumption of an exact eigenstate warm start, we prove a one-step \emph{propagation} inequality:
\[
\Std_\psi(\lambda^+)\le\Std_\psi(\lambda)+|\delta\lambda|\,\Std_\psi(H_\text{f}-H_\text{i}),
\]
based on a covariance bound (Cauchy--Schwarz).
This yields a directly checkable sufficient condition for preserving the unique branch at $\lambda^+$ from measurements at $\lambda$.
This is done in Lemmas~\ref{lem:covariance_bound}--\ref{lem:variance_propagation} and Corollary~\ref{cor:verification_condition}.

\item \textbf{Coupling verification to PŁ tracking.}
Finally, we combine the verification certificate with the PŁ-based contraction guarantee from the optimization analysis. If the warm start is certified to lie on the ground branch at $\lambda^+$ and the PŁ region drifts continuously, then (i) the same parameters remain inside the PŁ region at $\lambda^+$ and (ii) subsequent gradient steps contract toward $\bth^\star(\lambda^+)$.
Starting from the exact ground state at $\lambda=0$, repeated application gives the ground branch at every step.
This yields a \emph{self-verified} adiabatic update rule.
This combination is formalized in \tref{thm:verification_pl_tracking}.

\end{enumerate}

\subsection{Lemmas and corollaries}

\begin{lemma}[Standard deviation identifies a unique eigen-branch]
\label{lem:variance_branch}
For any $\lambda$ and any normalized state $\ket{\psi}$, there exists an index $j$ such that
\begin{equation}
\big| E_\psi(\lambda) - E_j(\lambda) \big|
\le \Std_\psi(\lambda) .
\end{equation}
If the tracked branch \(j_\star\) is one such index and \(\Std_\psi(\lambda)<\deltaminj/2\), then \(j_\star\) is the unique such index.
\end{lemma}

\begin{proof}
Let $p_j = \bra{\psi} \Pi_j(\lambda) \ket{\psi}$.
Then
\[
\Std_\psi(\lambda)^2
=
\sum_{j\ge 0} p_j \big(E_j(\lambda) - E_\psi(\lambda)\big)^2 .
\]
If $|E_j - E_\psi| > \Std_\psi$ for all $j$, then
\[
\Std_\psi^2
>
\sum_j p_j \Std_\psi^2
= \Std_\psi^2,
\]
which is a contradiction.
Hence, at least one index $j$ satisfies the bound.
If the tracked index \(j_\star\) and another index \(k\neq j_\star\) both satisfy the bound, then
\[
|E_{j_\star} - E_k|
\le |E_{j_\star} - E_\psi| + |E_k - E_\psi|
\le 2\Std_\psi .
\]
If \(\Std_\psi<\deltaminj/2\), this contradicts Assumption~\ref{ass:global_gap_ver}.
\end{proof}

\begin{lemma}[Standard deviation-to-fidelity bound for the identified branch]
\label{lem:var_to_fid_branch}
Fix $\lambda$ and a normalized state $\ket{\psi}$.
Assume $\Std_\psi(\lambda) < \deltaminj/2$ and let $j_\star$ be the unique tracked index from \lref{lem:variance_branch}, i.e.
\[
|E_\psi(\lambda) - E_{j_\star}(\lambda)| \le \Std_\psi(\lambda) .
\]
Then
\begin{equation}
1 - \bra{\psi}\Pi_{j_\star}(\lambda)\ket{\psi}
\le
\frac{\Std_\psi(\lambda)^2}
{\big(\deltaminj - \Std_\psi(\lambda)\big)^2}.
\label{eq:branch_fidelity}
\end{equation}
\end{lemma}

\begin{proof}
Expand $\ket{\psi}$ in the eigenbasis of $H(\lambda)$: $\ket{\psi} = \sum_j \alpha_j \ket{E_j(\lambda)}$ with $p_j = |\alpha_j|^2=\bra{\psi}\Pi_j\ket{\psi}$.
Then
\[
\Std_\psi(\lambda)^2
=
\sum_j p_j \big(E_j(\lambda) - E_\psi(\lambda)\big)^2 .
\]
Split the sum into the identified branch $j_*$ and its complement:
\[
\Std_\psi(\lambda)^2
=
p_{j_\star}\big(E_{j_\star}-E_\psi\big)^2
+
\sum_{j\neq j_\star} p_j\big(E_j-E_\psi\big)^2
\ge
\sum_{j\neq j_\star} p_j\big(E_j-E_\psi\big)^2,
\]
where we used the fact that each term is nonnegative.
For any $j\neq j_\star$,
\[
|E_j - E_\psi|
\ge
|E_j - E_{j_\star}| - |E_{j_\star} - E_\psi|
\ge
\deltaminj - \Std_\psi(\lambda),
\]
where Assumption~\ref{ass:global_gap_ver} gives \(|E_j-E_{j_\star}|\geq\deltaminj\), while the hypothesis that \(j_\star\) is the identified index gives \(|E_{j_\star}-E_\psi|\leq\Std_\psi(\lambda)\), and hence \(-|E_{j_\star}-E_\psi|\geq-\Std_\psi(\lambda)\).
Therefore,
\[
\Std_\psi(\lambda)^2
\ge
\sum_{j\neq j_\star} p_j
\big(\deltaminj - \Std_\psi(\lambda)\big)^2
=
\big(1 - p_{j_\star}\big)
\big(\deltaminj - \Std_\psi(\lambda)\big)^2.
\]
Rearranging yields~\eqref{eq:branch_fidelity}.
\end{proof}

\begin{corollary}[Ground-state fidelity once the branch is known]
\label{cor:ground_fidelity}
Under Assumption~\ref{ass:global_gap_ver}, suppose that the unique index $j_\star$ in \lref{lem:variance_branch} equals $0$, i.e., we consider the ground state.
If $\Std_\psi(\lambda) \le \deltamin\,/4$, then
\[
\bra{\psi}\Pi_0(\lambda)\ket{\psi}
\ge
\frac{8}{9}.
\]
\end{corollary}

\begin{proof}
Apply \lref{lem:var_to_fid_branch} with $j_\star = 0$.
If $\Std_\psi \le \deltamin\,/4$, then $\deltamin\, - \Std_\psi \ge 3\deltamin\,/4$, giving the stated bound.
\end{proof}

The quarter-gap threshold gives the stated \(8/9\) fidelity; any smaller-than-half threshold gives the corresponding bound in \eref{eq:branch_fidelity}.

\begin{lemma}[Exact warm-start standard deviation]
\label{lem:warm_start_exact}
Let $\ket{0(\lambda)}$ denote the ground state of $H(\lambda)$.
Then for any $\delta\lambda$,
\begin{equation}
\operatorname{\Std}_{0(\lambda)}\!\big(H(\lambda+\delta\lambda)\big)^2
=
(\delta\lambda)^2\,
\operatorname{\Std}_{0(\lambda)}\!\big(H_{\mathrm f}-H_{\mathrm i}\big)^2.
\end{equation}
\end{lemma}

\begin{proof}
For the linear schedule,
\[
H(\lambda+\delta\lambda)
=H(\lambda)+\delta\lambda(H_{\mathrm f}-H_{\mathrm i}).
\]
Since \(H(\lambda)\ket{0(\lambda)}=E_0(\lambda)\ket{0(\lambda)}\), the first moment is
\begin{align*}
&\bra{0(\lambda)}H(\lambda+\delta\lambda)\ket{0(\lambda)}\\
&\qquad=E_0(\lambda)
+\delta\lambda\,
\bra{0(\lambda)}(H_{\mathrm f}-H_{\mathrm i})\ket{0(\lambda)}.
\end{align*}
The second moment is
\begin{align*}
&\bra{0(\lambda)}H(\lambda+\delta\lambda)^2\ket{0(\lambda)}\\
&\quad=E_0(\lambda)^2
+2\delta\lambda\,E_0(\lambda)
\bra{0(\lambda)}(H_{\mathrm f}-H_{\mathrm i})\ket{0(\lambda)}\\
&\qquad+(\delta\lambda)^2
\bra{0(\lambda)}(H_{\mathrm f}-H_{\mathrm i})^2\ket{0(\lambda)}.
\end{align*}
Here both cross terms reduce to the same expression because \(H(\lambda)\ket{0(\lambda)}=E_0(\lambda)\ket{0(\lambda)}\) and, since \(H(\lambda)\) is Hermitian, \(\bra{0(\lambda)}H(\lambda)=E_0(\lambda)\bra{0(\lambda)}\).
Subtracting the square of the first moment from the second moment gives
\begin{align*}
&\operatorname{\Std}_{0(\lambda)}
\!\big(H(\lambda+\delta\lambda)\big)^2\\
&\quad=(\delta\lambda)^2\left[
\bra{0(\lambda)}(H_{\mathrm f}-H_{\mathrm i})^2\ket{0(\lambda)}
-\left(
\bra{0(\lambda)}(H_{\mathrm f}-H_{\mathrm i})\ket{0(\lambda)}
\right)^2
\right]\\
&\quad=(\delta\lambda)^2\,
\operatorname{\Std}_{0(\lambda)}(H_{\mathrm f}-H_{\mathrm i})^2.
\end{align*}
This proves the claim.
\end{proof}

\begin{lemma}[Verification step-size bound]
\label{lem:vericication_step}
If
\[
|\delta\lambda| < \frac{\deltamin\,} {2\operatorname{\Std}_{0(\lambda)}(H_{\mathrm f}-H_{\mathrm i})},
\]
then
\[
\operatorname{\Std}_{0(\lambda)}\!\big(H(\lambda+\delta\lambda)\big)<\frac{\deltamin\,}{2}.
\]
Consequently, the warm-start state $\ket{0(\lambda)}$ remains associated with the unique ground-state branch of \(H(\lambda+\delta\lambda)\).
\end{lemma}

\begin{proof}
The standard deviation bound follows directly from \lref{lem:warm_start_exact}.
The same strict half-gap bound holds at every intermediate point.
Since the associated branch is the ground branch at \(\delta\lambda=0\), continuity and uniqueness from \lref{lem:variance_branch} preserve that branch throughout the step, by the intermediate-value argument given in the proof of \cref{cor:verification_condition} below.
\end{proof}

\begin{lemma}[Covariance bound]
\label{lem:covariance_bound}
For any Hermitian operators $A,B$ and any normalized state $\ket{\psi}$,
\[
\big|\mathrm{Cov}_\psi(A,B)\big|=
\big|
\bra{\psi} AB \ket{\psi}
-
\bra{\psi} A \ket{\psi}
\bra{\psi} B \ket{\psi}
\big|
\le
\operatorname{\Std}_\psi(A)\operatorname{\Std}_\psi(B) .
\]
\end{lemma}

\begin{proof}
Let
\[
\bar A = \bra{\psi}A\ket{\psi},\qquad \bar B = \bra{\psi}B\ket{\psi},
\]
and define the centered (Hermitian) operators
\[
\widetilde A = A-\bar A\,\mathbb I,\qquad \widetilde B = B-\bar B\,\mathbb I.
\]
Then
\begin{align}
\bra{\psi}AB\ket{\psi}-\bra{\psi}A\ket{\psi}\bra{\psi}B\ket{\psi}
&=
\bra{\psi}(A-\bar A\mathbb I)(B-\bar B\mathbb I)\ket{\psi} \notag\\
&=
\bra{\psi}\widetilde A\,\widetilde B\ket{\psi}.
\label{eq:cov_centered}
\end{align}
Now set $\ket{u}=\widetilde A\ket{\psi}$ and $\ket{v}=\widetilde B\ket{\psi}$.
By the Cauchy--Schwarz inequality,
\[
\big|\bra{\psi}\widetilde A\,\widetilde B\ket{\psi}\big|
=
|\langle u|v\rangle|
\le
\|u\|\,\|v\|.
\]
Moreover,
\[
\|u\|^2
=
\bra{\psi}\widetilde A^2\ket{\psi}
=
\bra{\psi}(A-\bar A\mathbb I)^2\ket{\psi}
=
\operatorname{\Std}_\psi(A)^2,
\]
and similarly $\|v\|^2=\operatorname{\Std}_\psi(B)^2$.
Combining these identities with \eqref{eq:cov_centered} yields
\[
\big|
\bra{\psi} AB \ket{\psi}
-
\bra{\psi} A \ket{\psi}
\bra{\psi} B \ket{\psi}
\big|
\le
{\operatorname{\Std}_\psi(A)\operatorname{\Std}_\psi(B)},
\]
which proves the claim.
\end{proof}

\begin{lemma}[One-step standard deviation propagation]
\label{lem:variance_propagation}
For any state $\ket{\psi}$ and any $\lambda,\delta\lambda$,
\[
\Std_\psi(\lambda+\delta\lambda)
\le
\Std_\psi(\lambda)
+
|\delta\lambda|
\operatorname{\Std}_\psi(H_{\mathrm f}-H_{\mathrm i}).
\]
\end{lemma}

\begin{proof}
Write
\[
H(\lambda+\delta\lambda)=H(\lambda)+\delta\lambda \mathcal{D},
\qquad
\mathcal{D}=H_{\mathrm f}-H_{\mathrm i}.
\]
Denote expectation values with respect to $\ket{\psi}$ by $\langle \cdot \rangle = \bra{\psi}(\cdot)\ket{\psi}$.
Then
\[
\Std_\psi(\lambda+\delta\lambda)^2
=
\big\langle (H+\delta\lambda \mathcal{D})^2 \big\rangle
-
\big\langle H+\delta\lambda \mathcal{D} \big\rangle^2 .
\]
Expanding and regrouping terms yields
\begin{align}
\Std_\psi(\lambda+\delta\lambda)^2
&=
\underbrace{\big(\langle H^2\rangle-\langle H\rangle^2\big)}
_{=\,\Std_\psi(\lambda)^2}
+\delta\lambda
\big(\langle H\mathcal{D}+\mathcal{D}H\rangle
-2\langle H\rangle\langle \mathcal{D}\rangle\big)
+(\delta\lambda)^2
\big(\langle \mathcal{D}^2\rangle-\langle \mathcal{D}\rangle^2\big) \notag\\
&=
\Std_\psi(\lambda)^2
+2\delta\lambda\,\operatorname{Re}\mathrm{Cov}_\psi(H,\mathcal{D})
+(\delta\lambda)^2\,\operatorname{\Std}_\psi(\mathcal{D})^2.
\label{eq:var_expanded}
\end{align}
By Lemma~\ref{lem:covariance_bound},
\[
\big|\operatorname{Re}\mathrm{Cov}_\psi(H,\mathcal{D})\big|
\le
\big|\mathrm{Cov}_\psi(H,\mathcal{D})\big|
\le
{\operatorname{\Std}_\psi(H)\operatorname{\Std}_\psi(\mathcal{D})}
=
{\Std_\psi(\lambda)\operatorname{\Std}_\psi(\mathcal{D})}.
\]
Inserting this bound into~\eqref{eq:var_expanded} gives
\[
\Std_\psi(\lambda+\delta\lambda)^2
\le
\Std_\psi(\lambda)^2
+2|\delta\lambda|
{\Std_\psi(\lambda)\operatorname{\Std}_\psi(\mathcal{D})}
+(\delta\lambda)^2\operatorname{\Std}_\psi(\mathcal{D})^2
=
\big(
{\Std_\psi(\lambda)}
+
|\delta\lambda|{\operatorname{\Std}_\psi(\mathcal{D})}
\big)^2.
\]
Taking square roots on both sides yields
\[
{\Std_\psi(\lambda+\delta\lambda)}
\le
{\Std_\psi(\lambda)}
+
|\delta\lambda|{\operatorname{\Std}_\psi(H_{\mathrm f}-H_{\mathrm i})},
\]
which proves the claim.
\end{proof}

The propagation bound controls only the \emph{size} of the standard deviation at \(\lambda+\delta\lambda\).
By \lref{lem:variance_branch}, a small standard deviation guarantees association with \emph{some} eigen-branch, but taken at a single point it does not identify \emph{which} one---it is equally consistent with an excited branch.
The following corollary closes this gap: a continuity argument along the interpolated path excludes such branch hopping, turning quantities measured at \(\lambda\) into a certificate for the ground branch at \(\lambda+\delta\lambda\).

\begin{corollary}[Verification condition]
\label{cor:verification_condition}
Assume that the unique eigen-branch associated with \(\ket{\psi}\) at \(\lambda\) is \(j_\star=0\).
If
\[
{\Std_\psi(\lambda)}
+
|\delta\lambda|
{\operatorname{\Std}_\psi(H_{\mathrm f}-H_{\mathrm i})}
<
\frac{\deltamin\,}{2},
\]
then \(\ket{\psi}\) remains associated with \(j_\star=0\) at \(\lambda+\delta\lambda\).
\end{corollary}

\begin{proof}
For $s\in[0,1]$ set \(\lambda_s=\lambda+s\delta\lambda\), and let \(\mu(s)\) and \(\sigma(s)\) denote the energy mean and standard deviation of the fixed state \(\ket\psi\) with respect to \(H(\lambda_s)\).
Applying \lref{lem:variance_propagation} with step \(s\delta\lambda\) and using the assumption gives, for every \(s\in[0,1]\),
\[
\sigma(s)
\le
{\Std_\psi(\lambda)}
+|\delta\lambda|{\operatorname{\Std}_\psi(H_{\mathrm f}-H_{\mathrm i})}
<
\frac{\deltamin}{2}.
\]
Consider the mean energy measured from the instantaneous ground level,
\[
f(s)=\mu(s)-E_0(\lambda_s)\geq0,
\]
a continuous function of \(s\) (the mean never lies below the ground energy).
By \lref{lem:variance_branch}, at every \(s\) some eigenvalue of \(H(\lambda_s)\) lies within \(\sigma(s)\) of \(\mu(s)\).
If that eigenvalue is the ground level, then \(f(s)\leq\sigma(s)<\deltamin/2\); if it is an excited level \(E_j\), then \(\mu(s)\geq E_j(\lambda_s)-\sigma(s)\) and Assumption~\ref{ass:global_gap_ver} gives \(f(s)\geq\deltamin-\sigma(s)>\deltamin/2\).
Thus, because the gap stays open, \(f\) can never take the value \(\deltamin/2\).
Since \(f(0)\leq\sigma(0)<\deltamin/2\) by the assumed association at \(\lambda\), \(f\) starts below this value.
If \(f(s_1)>\deltamin/2\) held for some \(s_1\), the intermediate value theorem would force \(f(s_2)=\deltamin/2\) at some intermediate \(s_2\in(0,s_1)\), which is impossible.
Hence \(f(s)<\deltamin/2\) for all \(s\), so the excited alternative never occurs and \(f(1)\leq\sigma(1)\), i.e., \(\ket\psi\) is associated with the ground branch at \(\lambda+\delta\lambda\); by \lref{lem:variance_branch} with \(\sigma(1)<\deltamin/2\), this branch is unique.
\end{proof}

\begin{remark}[Exact-quadratic certified step]
\label{rem:maximal_step}
The step rule derived from \lref{lem:variance_propagation} uses only \(\Std_\psi(\lambda)\) and \(\Std_\psi(H_{\mathrm f}-H_{\mathrm i})\), at the price of the Cauchy--Schwarz relaxation.
Alternatively, the exact expansion~\eqref{eq:var_expanded} shows that \(\Std_\psi(\lambda+\delta\lambda)^2\) is a convex quadratic in \(\delta\lambda\) whose three coefficients---\(\Var_\psi(H(\lambda))\), \(\operatorname{Re}\mathrm{Cov}_\psi(H,D)\) with \(D=H_{\mathrm f}-H_{\mathrm i}\), measurable via the Hermitian observable \((HD+DH)/2\), and \(\Var_\psi(D)\)---can all be estimated on the current state.
Taking \(\delta\lambda\) as a fraction \(\kappa\in(0,1)\) of the largest value for which \(\Std_\psi(\lambda+\delta\lambda)<\deltamin/2\) therefore yields the largest admissible step up to the safety factor.
By convexity, the standard deviation along the whole segment is bounded by its values at the two endpoints, so the branch-preservation proof of \cref{cor:verification_condition} applies verbatim, with its first step supplied by convexity instead of the propagation bound.
For forward steps, the Cauchy--Schwarz rule is recovered in the worst case \(\operatorname{Re}\mathrm{Cov}_\psi(H,D) =\Std_\psi(\lambda)\,\Std_\psi(D)\).
\end{remark}

\begin{theorem}[Certified evolution from PŁ entry and branch verification]
\label{thm:verification_pl_tracking}
Fix $\lambda$ and $\lambda^+=\lambda+\delta\lambda$.
Let $\bth$ be parameters at $\lambda$ with prepared state $\ket{\psi}=\ket{\psi(\bth)}$.
Define
\[
\delta\lambda_{\rm A} = \frac{\gamma^{2}\deltamin^{2}} {16\,\|H\|_{\mathrm{op}}\, M^{2}\,\|H_\text{f}-H_\text{i}\|_{\mathrm{op}}}, 
\qquad 
\delta\lambda_{\rm V} =\frac{\deltamin/2-\Std_\psi(\lambda)} {\Std_\psi(H_\text{f}-H_\text{i})}.
\]
Assume:
\begin{enumerate}
\item the unique eigen-branch associated with \(\ket{\psi}\) at \(\lambda\) is \(j_\star=0\);
\item \(\Std_\psi(\lambda)\leq\deltamin/4\);
\item \(\|\bth-\bth^\star(\lambda)\|_2\leq r_{\rm PL}/2\);
\item $|\delta\lambda|\leq\delta\lambda_{\rm A}$ and $|\delta\lambda|<\delta\lambda_{\rm V}$.
\item the number of vanilla gradient-descent steps satisfies \( K\geq L(\gamma\deltamin)^{-1}\ln2 \).
\end{enumerate}
Then:
\begin{enumerate}
\item $\ket{\psi}$ remains associated with the unique ground-state branch of $H(\lambda^+)$;
\item the same parameters $\bth$ lie inside the PŁ region at $\lambda^+$;
\item the \(K\) vanilla gradient-descent steps with \(\eta=1/L\) remain in the PŁ region, do not increase the energy, and return an output satisfying \( \|\bth_{\rm out}-\bth^\star(\lambda^+)\|_2\leq r_{\rm PL}/2 \);
\item if after optimization at $\lambda^+$ the algorithm enforces \(\Std_{\psi(\bth_{\mathrm{out}})}(\lambda^+)\leq\deltamin/4\), then the output state remains on the same eigen-branch with overlap at least $8/9$.
\end{enumerate}
\end{theorem}

\begin{proof}
Because $|\delta\lambda|<\delta\lambda_{\rm V}$, the propagated standard-deviation bound
\[
\Std_\psi(\lambda^+)
\le \Std_\psi(\lambda)
+|\delta\lambda|\Std_\psi(H_\text{f}-H_\text{i})
\]
implies $\Std_\psi(\lambda^+)<\deltamin/2$, so \cref{cor:verification_condition} yields (1).

Because $|\delta\lambda|\leq\delta\lambda_{\rm A}$, the drift bound and the triangle inequality give
\[
\|\bth-\bth^\star(\lambda^+)\|_2
\leq
\|\bth-\bth^\star(\lambda)\|_2
+D|\delta\lambda|
\leq r_{\rm PL}.
\]
This proves (2).
By \lref{lem: warm-start}, the iterates remain in this region, and the stated condition on \(K\) contracts the terminal distance by at least \(1/2\), proving (3).

For (4), let \(\mu_{\rm in}\) and \(\mu_{\rm out}\) be the energies before and after optimization at \(\lambda^+\).
By (1), \(\mu_{\rm in}<E_0(\lambda^+)+\deltamin/2\), and energy descent gives \(\mu_{\rm out}\leq\mu_{\rm in}\).
An excited associated branch would instead imply
\[
\mu_{\rm out}\geq E_1(\lambda^+)
-\Std_{\psi(\bth_{\rm out})}(\lambda^+)
\geq E_0(\lambda^+)+\frac{3\deltamin}{4}
>E_0(\lambda^+)+\frac{\deltamin}{2},
\]
a contradiction.
Thus, the output remains on the ground branch, and Lemma~\ref{lem:var_to_fid_branch} gives (4).
\end{proof}

At \(t=0\), exact initialization gives \(j_\star=0\); applying Theorem~\ref{thm:verification_pl_tracking} successively proves the claim for every accepted output.

\begin{remark}[Adaptive schedule without trainability constants]
\label{rem:adaptive_schedule}
The algorithmic inputs that involve trainability constants---the step size \(\delta\lambda_{\rm A}\) and the iteration number \(K\)---may be replaced by arbitrary initial guesses \(\delta\lambda_0\) and \(K_0\): whenever the output test \(\Std_{\psi}(\lambda)\leq\Delta_c/4\) (with finite shots, any acceptance test of \sref{sec:measurement_cost}) fails to pass after the current budget of gradient-descent steps, restart the evolution from \(\lambda=0\) with the step cap halved and the budget doubled (with finite shots, also all shot counts).
Soundness is unaffected: within every run, each accepted slice still satisfies the strict step-size condition and the output test, so the branch-preservation and output-test arguments---items (1) and (4) of Theorem~\ref{thm:verification_pl_tracking}, whose proofs use only these measured conditions together with energy descent at \(\eta=1/L\)---certify every accepted state, and any run that reaches \(\lambda=1\) outputs a certified ground state.
With finite shots, energy descent is replaced by the measured energy test \eref{eq:finite_shot_energy_check}, so every accepted state remains certified.
Each run uses the failure budgets \(\delta/(R_{\max}+1)\) and \(\delta_{\rm cert}/(R_{\max}+1)\) of \sref{sec:measurement_cost}, where \(R_{\max}\) is a prescribed cap on the number of restarts, after which the protocol stops without output; hence all certificates of all runs are valid with probability at least \(1-\delta_{\rm cert}\).
Termination follows from the guarantees themselves.
In the exact case, with \(W=\max_\lambda[E_{\max}(\lambda)-E_0(\lambda)]\) the spectral width,
\[
\Std_{\psi(\bth)}(\lambda)^2
\leq\|(H-E_0)\psi(\bth)\|^2
\leq W^2\,\|\psi(\bth)-\psi_0\|^2
\leq W^2M\,\|\bth-\bth^\star\|_2^2 ,
\]
so the linear contraction of \lref{lem: warm-start} both maintains the tracking induction and drives the measured standard deviation below \(\Delta_c/4\) within \(K^\ast=\big\lceil\tfrac{L}{\gamma\deltamin} \ln\max\big\{2,\tfrac{4W\sqrt M\,r_{\rm PL}}{\Delta_c}\big\}\big\rceil\) steps from anywhere in the tracking region.
Hence, under Assumptions~1--3 of the main text, a run with step cap at most \(\delta\lambda_{\rm A}\) and budget at least \(K^\ast\) completes without any test failure, and at most \(\max\{0,\lceil\log_2(\delta\lambda_0/\delta\lambda_{\rm A})\rceil\} +\max\{0,\lceil\log_2(K^\ast/K_0)\rceil\}\) restarts occur---logarithmically many in the quality of the initial guesses.
The constants \(\gamma\), \(\deltamin\), and the \(\mu_{\mathrm{opt}}\) of \sref{sec:epsilon_representability} therefore enter only the complexity analysis, never the algorithm's inputs.
With finite shots, \(K\) and the shot count \(S\) must also satisfy \eref{eq:terminal_radius_conditions} and \eref{eq:margin_conditions}, with \(\xi_{\max}\) given by \eref{eq:xi_uniform_bound}, and the certificate shot counts those of \sref{subsec:finite_shot_certificate}; doubling them at restarts ensures this after logarithmically many additional restarts, after which a run completes with probability at least \(1-(\delta+\delta_{\rm cert})/(R_{\max}+1)\), provided \(R_{\max}\) is at least the number of restarts required.
Since every restart at least doubles the a priori cost bound of a run, all runs together cost at most twice that of the final run.
In practice, restoring the most recent certified pair \((\lambda,\bth)\) is more economical than a full restart; the restart form is stated because it makes the restart count immediate.
\end{remark}

\section{Extension to $\epsilon$-exact representability}
\label{sec:epsilon_representability}

This section generalizes the representability and nondegeneracy assumptions of the main text (Assumptions~2 and~3):
\aref{ass:epsilon_representability} replaces exact representability by an energy error \(\epsilon\), and \aref{ass:optimum_branch} replaces the exact minimum by a nondegenerate branch of variational optima.
Both are stated first, followed by their tracking and certification consequences.

\begin{assumption}[$\epsilon$-exact representability]
\label{ass:epsilon_representability}
There exists a constant $\epsilon\ge 0$ such that for all $\lambda\in[0,1]$ there exist parameters $\bar\bth(\lambda)$ satisfying
\begin{equation}
E_\lambda(\bar\bth(\lambda))
\le E_0(\lambda)+\epsilon,
\end{equation}
where $E_0(\lambda)$ denotes the exact ground-state energy.
\end{assumption}

For each \(\lambda\), fix a variational optimum, its energy, and its state,
\begin{equation}
\bth^{\mathrm{opt}}(\lambda)
\in \arg\min_{\bth} E_\lambda(\bth), \qquad
E_\lambda^{\mathrm{opt}} = E_\lambda(\bth^{\mathrm{opt}}(\lambda)), \qquad
\ket{\psi^{\mathrm{opt}}(\lambda)}=\ket{\psi(\bth^{\mathrm{opt}}(\lambda))}.
\end{equation}
Throughout this section, the superscript ``opt'' marks quantities of the optimal \emph{variational} state, while the exact ground state and its energy keep the symbols \(\ket{\psi_0(\lambda)}\) and \(E_0(\lambda)\).
In the exact case \(\epsilon=0\), the variational optimum reduces to the starred exact-ground-state parameters \(\bth^\star(\lambda)\) of the main text, and \(\ket{\psi^{\mathrm{opt}}}=\ket{\psi(\bth^\star)}=\ket{\psi_0}\).
The variational principle and Assumption~\ref{ass:epsilon_representability} then give
\begin{equation}
0\leq e_\lambda:=
E_\lambda^{\mathrm{opt}}-E_0(\lambda)\leq\epsilon.
\end{equation}

A first consequence of \aref{ass:epsilon_representability} is a variance bound at the optimum.
Let
\[
W_\lambda=E_{\max}(\lambda)-E_0(\lambda),\qquad
W=\max_{\lambda\in[0,1]}W_\lambda,\qquad
X_\lambda=H(\lambda)-E_0(\lambda)I,
\]
where \(E_{\max}(\lambda)\) denotes the largest eigenvalue of \(H(\lambda)\).
Since \(0\preceq X_\lambda\preceq W_\lambda I\) and \(X_\lambda\) commutes with \(W_\lambda I-X_\lambda\), the product \(X_\lambda(W_\lambda I-X_\lambda)\succeq0\), i.e., \(X_\lambda^2\preceq W_\lambda X_\lambda\).
Let \(\langle\cdot\rangle_{\mathrm{opt}}\) denote expectation values in the optimal variational state \(\ket{\psi^{\mathrm{opt}}(\lambda)}\), and \(\Var_{\mathrm{opt}}\) and \(\mathrm{Cov}_{\mathrm{opt}}\) the corresponding variance and covariance.
The constant shift leaves the variance unchanged, so, using \(\langle X_\lambda\rangle_{\mathrm{opt}} =E_\lambda^{\mathrm{opt}}-E_0(\lambda)=e_\lambda\),
\[
\Var_{\mathrm{opt}}(H(\lambda))
=\Var_{\mathrm{opt}}(X_\lambda)
=\langle X_\lambda^{2}\rangle_{\mathrm{opt}}
-\langle X_\lambda\rangle_{\mathrm{opt}}^{2}
=\langle X_\lambda^{2}\rangle_{\mathrm{opt}}-e_\lambda^{2}.
\]
Taking the expectation value of \(X_\lambda^{2}\preceq W_\lambda X_\lambda\) gives \(\langle X_\lambda^{2}\rangle_{\mathrm{opt}}\leq W_\lambda e_\lambda\), and hence
\begin{equation}
\Var_{\mathrm{opt}}(H(\lambda))
\leq W_\lambda e_\lambda-e_\lambda^2
=e_\lambda(W_\lambda-e_\lambda)
\leq W\epsilon ,
\end{equation}
where the last inequality uses \(e_\lambda\leq\epsilon\) and \(W_\lambda-e_\lambda\leq W\).

Energy error alone does not imply local curvature or a smooth branch of variational optima.
Tracking, therefore, also requires the following generalization of Assumption~3 of the main text, formulated on the branch of variational optima.

\begin{assumption}[Nondegenerate optimum branch]
\label{ass:optimum_branch}
In the fixed coordinates, \(\bth^{\mathrm{opt}}(\lambda)\) is \(C^1\) and, uniformly in \(\lambda\),
\[
\nabla^2_\bth E_\lambda(\bth^{\mathrm{opt}}(\lambda))
\succeq2\mu_{\mathrm{opt}}I,
\]
with \(\mu_{\mathrm{opt}}>0\).
\end{assumption}

In the exact case \(\epsilon=0\), Assumptions~2 and~3 of the main text imply Assumption~\ref{ass:optimum_branch} with \(\bth^{\mathrm{opt}}=\bth^\star\), \(\psi^{\mathrm{opt}}=\psi_0\), and \(\mu_{\mathrm{opt}}=\gamma\deltamin\), by \lref{lem: hess lower bound}.

\subsection{Tracking}

By \lref{lem: hessian lipschitz}, the assumed curvature extends to the ball of radius \(\mu_{\mathrm{opt}}/L_H\) around \(\bth^{\mathrm{opt}}(\lambda)\), exactly as in the exact-representability case.
Define
\[
D_{\mathrm{opt}}
=\max_{\lambda\in[0,1]}
\|\dot{\bth}^{\mathrm{opt}}(\lambda)\|_2.
\]
Differentiating the stationarity condition along the branch, as in \lref{lem:drift-bound}, gives \(D_{\mathrm{opt}}\leq\sqrt M\,\|\partial_\lambda H\|_{\rm op}/\mu_{\mathrm{opt}}\), so the explicit constants of Theorem~1 of the main text hold with \(\gamma\deltamin\) replaced by \(\mu_{\mathrm{opt}}\).
If \(\|\bth_0-\bth^{\mathrm{opt}}(0)\|_2\leq r_{\rm PL}/2\), then the tracking induction applies with
\[
r_{\rm PL}=\frac{\mu_{\mathrm{opt}}}{L_H},\qquad
\delta\lambda\leq\frac{r_{\rm PL}}{2D_{\mathrm{opt}}},
\qquad
K\geq\frac{L}{\mu_{\mathrm{opt}}}\ln2.
\]
For small \(\epsilon\), the curvature hypothesis of \aref{ass:optimum_branch} need not be assumed: it follows from the direct analog of the metric-nondegeneracy condition of Assumption~3 of the main text, now imposed on the optimum branch, and the exact-case curvature is recovered up to a factor of two.

\begin{lemma}[Curvature from the metric and the gap]
\label{lem: curvature small epsilon}
Let \(d\) be the number of retained coordinates and suppose that, in the fixed coordinates and uniformly in \(\lambda\), \(\bg(\bth^{\mathrm{opt}}(\lambda))\succeq\gamma I\) with \(\gamma>0\).
Under Assumption~1 of the main text and Assumption~\ref{ass:epsilon_representability}, if
\begin{equation}
\epsilon\leq\frac{\gamma^2\deltamin^2}{1600\,d^2\,W},
\label{eq:small_epsilon_condition}
\end{equation}
then, uniformly in \(\lambda\),
\begin{equation}
\nabla^2_\bth E_\lambda(\bth^{\mathrm{opt}}(\lambda))
\succeq\gamma\deltamin\,I ,
\label{eq:curvature_small_epsilon}
\end{equation}
i.e., the curvature bound of Assumption~\ref{ass:optimum_branch} holds with \(\mu_{\mathrm{opt}}=\gamma\deltamin/2\).
\end{lemma}

\begin{proof}
Fix \(\lambda\) and suppress it where convenient: write \(\psi^{\mathrm{opt}}\) for the optimal variational state \(\psi^{\mathrm{opt}}(\lambda)\), \(E=E_\lambda^{\mathrm{opt}}=E_0+e_\lambda\) for its energy, \(\psi_0\) and \(E_0\) for the exact ground state and its energy, and \(\Delta=\Delta(\lambda)\geq\deltamin\) for the gap.
All parameter derivatives \(\partial_\mu\psi\) of the ansatz state are evaluated at \(\bth^{\mathrm{opt}}(\lambda)\).
Let \(v\in\mathbb R^d\) be an arbitrary unit vector in the fixed coordinates; the goal is the scalar bound \(v^\top\nabla^2_\bth E_\lambda\,v\geq\gamma\deltamin\) at \(\bth^{\mathrm{opt}}(\lambda)\).
We abbreviate the first and second directional derivatives of the state along \(v\) as
\[
\phi=\sum_\mu v_\mu\partial_\mu\psi,
\qquad
\partial_v^2\psi=\sum_{\mu,\nu}v_\mu v_\nu\partial_\mu\partial_\nu\psi ;
\]
all parameter indices are contracted with \(v\), so both \(\phi\) and \(\partial_v^2\psi\) are single vectors in the Hilbert space.

\emph{Step 1: elementary norm bounds.}
Each derivative of the Pauli-rotation ansatz inserts a conjugated Pauli factor \(\widetilde P_\mu\) with \(\|\widetilde P_\mu\|_{\rm op}=1\): \(\partial_\mu\psi=-i\widetilde P_\mu\psi^{\mathrm{opt}}\), and similarly for mixed second derivatives.
Hence \(\|\partial_\mu\psi\|\leq1\) and \(\|\partial_\mu\partial_\nu\psi\|\leq1\), so the triangle inequality and the Cauchy--Schwarz bound \(\sum_\mu|v_\mu|\leq\sqrt d\,\|v\|_2=\sqrt d\) give
\[
\|\phi\|\leq\sum_\mu|v_\mu|\,\|\partial_\mu\psi\|\leq\sqrt d,
\qquad
\|\partial_v^2\psi\|
\leq\sum_{\mu,\nu}|v_\mu||v_\nu|\,\|\partial_\mu\partial_\nu\psi\|
\leq\Big(\sum_\mu|v_\mu|\Big)^2\leq d .
\]
Moreover, \(g_{\mu\nu}=\operatorname{Re}\mathrm{Cov}_{\mathrm{opt}}(\widetilde P_\mu,\widetilde P_\nu)\) satisfies \(|g_{\mu\nu}|\leq1\), hence \(\gamma\leq\|\bg\|_{\rm op}\leq d\).
Together with \(\deltamin\leq\Delta\leq W\), the hypothesis \eref{eq:small_epsilon_condition} therefore implies
\[
\epsilon
\leq\frac{\gamma^2}{d^2}\cdot\frac{\deltamin}{W}\cdot\frac{\deltamin}{1600}
\leq\frac{\deltamin}{1600} .
\]

\emph{Step 2: the Hessian splits into two terms.}
Differentiating \(E_\lambda(\bth)=\bra{\psi(\bth)}H\ket{\psi(\bth)}\) twice along \(v\), and subtracting \(E\) times the vanishing second derivative of the normalization \(\langle\psi(\bth)|\psi(\bth)\rangle=1\), gives
\[
v^\top\nabla^2_\bth E_\lambda\,v
=2\underbrace{\langle\phi|(H-E)|\phi\rangle}_{\text{curvature term}}
+2\underbrace{\operatorname{Re}
\langle\partial_v^2\psi|(H-E)|\psi^{\mathrm{opt}}\rangle}
_{\text{residual term}} .
\]
The plan: the curvature term reproduces the exact-case bound \(\gamma\Delta\) up to corrections of order \(\sqrt{W\epsilon}\), and the residual term is itself of order \(\sqrt{W\epsilon}\), because \((H-E)\psi^{\mathrm{opt}}\) would vanish for an exact eigenstate.

\emph{Step 3: the residual term.}
Since \(E=\langle H\rangle_{\mathrm{opt}}\), the norm of the residual vector is the energy standard deviation, \(\|(H-E)\psi^{\mathrm{opt}}\| =\sqrt{\Var_{\mathrm{opt}}(H(\lambda))}\).
Hence, by Cauchy--Schwarz,
\[
2\big|\operatorname{Re}
\langle\partial_v^2\psi|(H-E)|\psi^{\mathrm{opt}}\rangle\big|
\leq2\,\|\partial_v^2\psi\|\,\sqrt{\Var_{\mathrm{opt}}(H(\lambda))} .
\]

\emph{Step 4: the curvature term via the spectral decomposition.}
Let \(a_0=\langle\psi_0|\phi\rangle\).
In the eigenbasis of \(H\), the ground level contributes \((E_0-E)|a_0|^2=-e_\lambda|a_0|^2\), while every excited level satisfies \(E_j-E\geq\Delta-e_\lambda\geq0\) (positivity holds because \(e_\lambda\leq\epsilon\leq\deltamin/1600\)).
The excited part of \(\phi\) has weight \(\|\phi\|^2-|a_0|^2\), so
\[
\langle\phi|(H-E)|\phi\rangle
\geq-e_\lambda|a_0|^2+(\Delta-e_\lambda)\big(\|\phi\|^2-|a_0|^2\big)
=(\Delta-e_\lambda)\|\phi\|^2-\Delta|a_0|^2 .
\]

\emph{Step 5: replacing the exact ground state by the variational state.}
The metric controls the overlap of \(\phi\) with \(\psi^{\mathrm{opt}}\) (Step 6), not with \(\psi_0\), so we compare the two.
Let \(p=1-|\langle\psi_0|\psi^{\mathrm{opt}}\rangle|^2\) denote the excited weight of the optimal variational state, and let \(p_j\) be its populations in the eigenbasis of \(H\).
Every excited level lies at least \(\Delta\) above \(E_0\), so the energy error dominates the excited weight:
\[
e_\lambda=\langle H-E_0\rangle_{\mathrm{opt}}
=\sum_{j\geq1}p_j\,(E_j-E_0)
\geq\Delta\sum_{j\geq1}p_j=\Delta\,p ,
\qquad\text{i.e.}\quad
p\leq\frac{e_\lambda}{\Delta} .
\]
Write \(\psi^{\mathrm{opt}}=c_0\psi_0+\chi\) with \(\langle\psi_0|\chi\rangle=0\), \(|c_0|^2=1-p\), \(\|\chi\|^2=p\).
Then \(\bar c_0\,a_0 =\langle\psi^{\mathrm{opt}}|\phi\rangle-\langle\chi|\phi\rangle\), with \(|\langle\chi|\phi\rangle|\leq\sqrt p\,\|\phi\|\) by Cauchy--Schwarz.
Expanding the square and bounding only the cross term, with \(|\langle\psi^{\mathrm{opt}}|\phi\rangle|\leq\|\phi\|\),
\[
(1-p)|a_0|^2
=\big|\langle\psi^{\mathrm{opt}}|\phi\rangle
-\langle\chi|\phi\rangle\big|^2
\leq|\langle\psi^{\mathrm{opt}}|\phi\rangle|^2
+\big(2\sqrt p+p\big)\|\phi\|^2 .
\]

\emph{Step 6: the metric enters.}
Define the Hermitian tangent operator \(A=\sum_\mu v_\mu\widetilde P_\mu\), so that \(\phi=-iA\psi^{\mathrm{opt}}\).
Then \(\|\phi\|^{2}=\langle A^{2}\rangle_{\mathrm{opt}}\) and \(\langle\psi^{\mathrm{opt}}|\phi\rangle=-i\langle A\rangle_{\mathrm{opt}}\), while contracting \(g_{\mu\nu}=\operatorname{Re}\mathrm{Cov}_{\mathrm{opt}}(\widetilde P_\mu,\widetilde P_\nu)\) with the real unit vector \(v\) gives
\[
v^\top\bg\,v
=\mathrm{Cov}_{\mathrm{opt}}(A,A)
=\langle A^{2}\rangle_{\mathrm{opt}}-\langle A\rangle_{\mathrm{opt}}^{2}
=\|\phi\|^{2}-|\langle\psi^{\mathrm{opt}}|\phi\rangle|^{2} .
\]
In words: the metric measures the norm of the component of \(\phi\) orthogonal to the state itself---exactly the part that must pay the spectral gap.
Substituting \(|\langle\psi^{\mathrm{opt}}|\phi\rangle|^{2}=\|\phi\|^{2}-v^\top\bg\,v\) into the Step 5 bound, dividing it by \(1-p\), and inserting the result into Step 4, the leading \(\Delta\|\phi\|^{2}\) contributions cancel and only \(O(\sqrt p)\) errors survive:
\begin{align*}
\langle\phi|(H-E)|\phi\rangle
&\geq(\Delta-e_\lambda)\|\phi\|^{2}
-\frac{\Delta}{1-p}
\Big(\|\phi\|^{2}-v^\top\bg\,v+\big(2\sqrt p+p\big)\|\phi\|^{2}\Big)\\
&=\frac{\Delta}{1-p}\,v^\top\bg\,v
-\Big(e_\lambda+\frac{2\sqrt p+2p}{1-p}\,\Delta\Big)\|\phi\|^{2} .
\end{align*}
In the exact case (\(e_\lambda=p=0\)) this is \(\langle\phi|(H-E)|\phi\rangle\geq\Delta\,v^\top\bg\,v\), the bound of \lref{lem: hess lower bound}.

\emph{Step 7: assembling and bounding all constants.}
Inserting Steps 3 and 6 into the split of Step 2 yields the master inequality
\[
v^\top\nabla^2_\bth E_\lambda\,v
\geq\frac{2\Delta}{1-p}\,v^\top\bg\,v
-2\Big(e_\lambda+\frac{2\sqrt p+2p}{1-p}\,\Delta\Big)\|\phi\|^{2}
-2\,\|\partial_v^2\psi\|\sqrt{\Var_{\mathrm{opt}}(H(\lambda))} .
\]
Up to here, the only inequalities used are the spectral bound (Step 4), the single cross-term estimate (Step 5), and Cauchy--Schwarz (Step 3); every constant is exact.
All bounding now occurs in a single sweep.
From Step 1 and the variance bound above: \(\|\phi\|^{2}\leq d\), \(\|\partial_v^2\psi\|\leq d\), and \(\Var_{\mathrm{opt}}(H(\lambda))\leq W\epsilon\).
From Steps 5 and 1: \(p\leq e_\lambda/\Delta\leq\epsilon/\deltamin\leq1/1600\), so in particular \(1/(1-p)\leq2\) and \(p\leq\sqrt p\).
The good term is then bounded using \(1/(1-p)\geq1\), the hypothesis \(v^\top\bg\,v\geq\gamma\), and \(\Delta\geq\deltamin\),
\[
\frac{2\Delta}{1-p}\,v^\top\bg\,v\geq2\gamma\deltamin ,
\]
while both error coefficients reduce to the single scale \(\sqrt{W\epsilon}\),
\[
\frac{2\sqrt p+2p}{1-p}\,\Delta
\leq8\Delta\sqrt p
\leq8\sqrt{\Delta e_\lambda}
\leq8\sqrt{W\epsilon},
\qquad
e_\lambda\leq\epsilon\leq\sqrt{W\epsilon},
\]
using \(\Delta\sqrt p\leq\sqrt{\Delta e_\lambda}\) (from \(p\leq e_\lambda/\Delta\)) and \(\Delta\leq W\), \(e_\lambda\leq\epsilon\leq W\).
Altogether,
\[
v^\top\nabla^2_\bth E_\lambda\,v
\geq2\gamma\deltamin
-2(1+8)\,d\sqrt{W\epsilon}-2d\sqrt{W\epsilon}
=2\gamma\deltamin-20\,d\sqrt{W\epsilon} .
\]
A single budget fixes the constant of the lemma: the choice \(1600=(2\cdot20)^2\) in \eqref{eq:small_epsilon_condition} is equivalent to \(\sqrt{W\epsilon}\leq\gamma\deltamin/(40d)\), i.e. \(20d\sqrt{W\epsilon}\leq\gamma\deltamin/2\), so that
\[
v^\top\nabla^2_\bth E_\lambda\,v
\geq2\gamma\deltamin-\frac{\gamma\deltamin}{2}
=\frac{3}{2}\gamma\deltamin
\geq\gamma\deltamin ,
\]
which is \eqref{eq:curvature_small_epsilon}.
\end{proof}

Since \(d\leq M\), the main-text condition \(\epsilon\leq\gamma^2\deltamin^2/(1600\,M^2\,W)\) implies \eref{eq:small_epsilon_condition}.

\subsection{Certification}

By the variance bound above, sufficient conditions for \(\ket{\psi^{\mathrm{opt}}}\) to satisfy the half-gap branch test and the quarter-gap output test are, respectively,
\begin{equation}
\epsilon<\frac{\Delta_c^2}{4W},
\qquad
\epsilon<\frac{\Delta_c^2}{16W}.
\end{equation}
The gap also gives the direct fidelity bound
\begin{equation}
|\langle\psi_0(\lambda)|
\psi^{\mathrm{opt}}(\lambda)\rangle|^2
\geq1-\frac{e_\lambda}{\Delta_c}
\geq1-\frac{\epsilon}{\Delta_c}.
\end{equation}
Variance- or fidelity-based representability assumptions state these conditions directly; the energy-based form is natural for VQE.
The spectral verification theorem itself is unchanged.

\section{Numerical analysis details}
\label{sec:numerical_analysis}

In this section, we supply additional detail to the numerical experiments described and presented in the main text.
Specifically, we show the ansatz for the circuits simulated in this manuscript, and justify a numerical adaptation when computing $\dl_{\mathrm{A}}$ in accordance with the self-verifying AVQE algorithm.
Although the adiabatic tracking step size $\dl$ can be treated as a hyperparameter, to ensure we remain within the basins of the minima along the adiabatic path, $\dl$ has to satisfy $\dl = \min\{\dl_{\mathrm{A}},\dl_{\mathrm{V}},1-\lambda\}$ for all points along the path.
Therefore, ensuring both reasonable values and efficient computation of $\dl_{\mathrm{A}}$, as well as $\dl_{\mathrm{V}}$, is necessary for a practical implementation.

\begin{figure}
\centering
\includegraphics[width=0.45\linewidth]{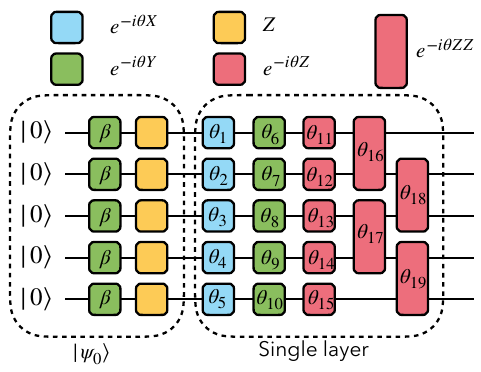}
\caption{\label{fig:S1}
Variational quantum circuit defining $\ket{\psi(\bth)}$ for one layer.
The initial state $\ket{\psi_0}$ is initialized as the ground state of the Hamiltonian at $\lambda=0$.
}
\end{figure}

Figure~\ref{fig:S1} reports the quantum circuit defining the parameterized quantum state $\ket{\psi(\bth)}$ used in the numerical experiments, for $n=5$ qubits and one layer.
The state is initialized in the ground state of $H(\lambda=0)$, by a single-qubit layer $\ket{\psi_0}=\bigotimes_{\ell=1}^n Z_\ell e^{-i \beta Y_\ell} \ket{0}$, where $\beta$ is computed analytically, since at $\lambda=0$ the exact ground state is a product state.

The formula for $\delta\lambda_{\mathrm{A}}$ derived in \sref{sec:daws1} is very useful to make theoretical claims on our protocol.
However, the upper bound provided can be loose. 
A better strategy to numerically probe our protocol is to avoid any bounding by computing quantities that are easily accessible numerically.
Thus, we rewrite the step size as
\begin{align}
\delta\lambda_A&\approx
\frac{\zeta_{\rm res}^2} {4L_H \norm{\nabla_\bth\expval{\partial_\lambda H(\lambda, \bth^\star)}}_2},
\end{align}
where \(\zeta_{\rm res}\) denotes the smallest resolved positive eigenvalue of the numerical Hessian under the heuristic protocol below.
It replaces the lower bound on the curvature, \(2\gamma\deltamin\) of Eq.~\eqref{eq:r_pl_2}; with this substitution, the displayed step size is the numerical analog of the warm-start condition of \sref{sec:daws1}.
This numerical estimate is not used in the proof of \texttt{Theorem 1}.
Finally, we need an efficient procedure to compute the Lipschitz constant $L_H$.
We use the definition of the local Lipschitz constant, but we approximate it by sampling around the minimum $\bth^\star$:
\begin{align}
L_H &\approx \max_{i} \frac{\norm{\mathcal{H}(\bth^\star)-\mathcal{H}(\bth^\star+\delta\bth^{(i)})}_{op}}{\norm{\delta\bth^{(i)}}_2} \\
\delta\bth^{(i)}_j &\sim \text{Uniform}(-r,r).
\end{align}

\begin{figure}
\centering
\begin{overpic}[width=0.45\linewidth]{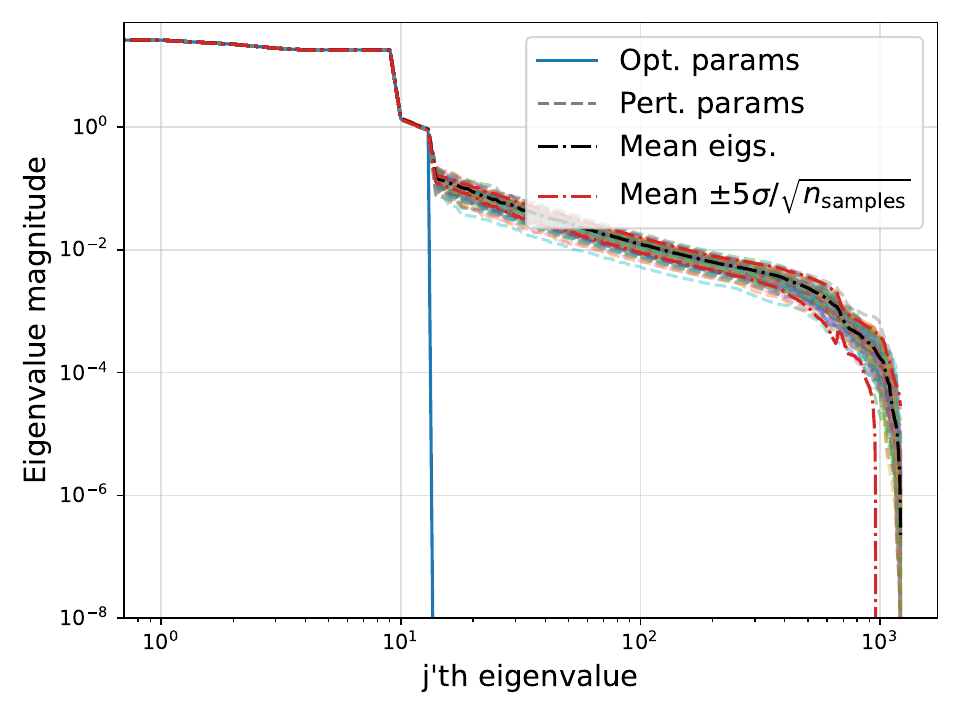}
\put(6,70){(a)}
\end{overpic}
\hfill
\begin{overpic}[width=0.45\linewidth]{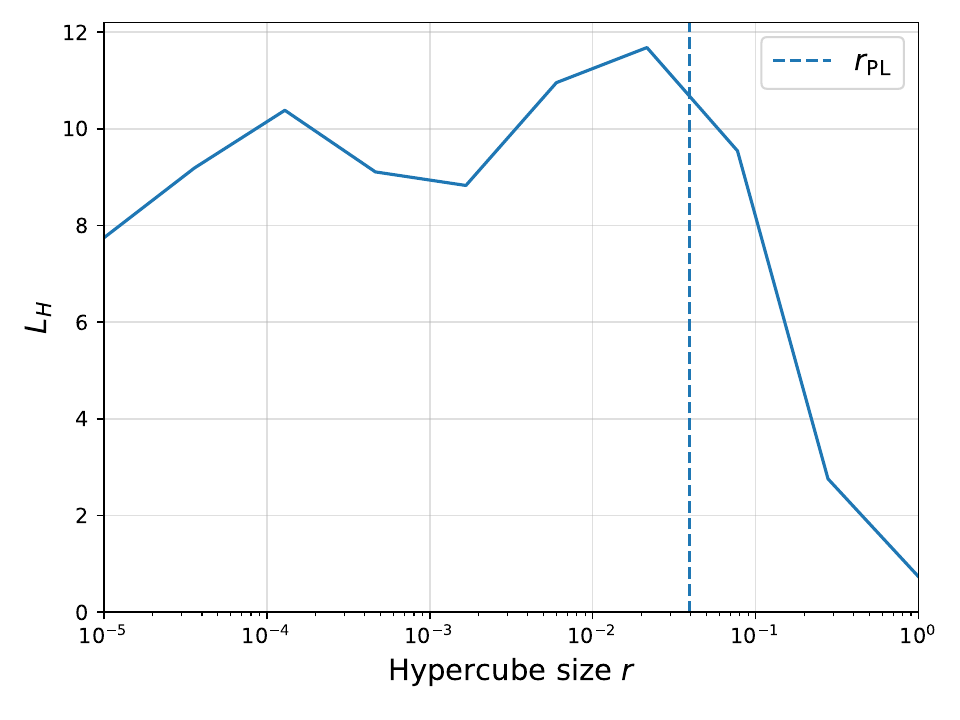}
\put(-2,70){(b)}
\end{overpic}
\caption{\label{fig:S2}
Parameter selection for computing $\dl_{\mathrm{A}}$ for 5 qubits.
Panel (a) shows the magnitudes of the Hessian eigenvalues at $\lambda = 0$ for the optimized and 100 samples of perturbed parameters.
Small signed negative values can occur because the numerical parameters are not exact stationary points; the magnitude plot is used only to display the
separation of scales.
We also show the mean result for 5 randomly selected samples with a sampling error of $\pm 5\sigma / \sqrt{n_{\mathrm{samples}}}$, where $\sigma$ is the standard deviation computed from $n_{\mathrm{samples}} = 5$.
Panel (b) shows how the numeric expression for $L_H$ varies with the size of the perturbation in the parameters, $\delta \theta \sim \mathrm{Uniform}(-r, r)$, using 5 samples.
}
\end{figure}

For the adjusted formula of the Lipschitz constant $L_H$, we require computing the Hessian with parameters randomly perturbed from their optimized value, with the perturbed values sampled from the uniform distribution $\delta \theta \sim \text{Uniform}(-r, r)$.
The numerical calculations use the full parameterization, whereas the fixed nonredundant coordinates are an assumption of Theorem~1.
The perturbations are therefore sampled in the full parameter space.
Redundant directions produce near-zero Hessian modes, which the protocol below separates from the physical spectrum.
Figure~\ref{fig:S2} (a) shows the spectra of eigenvalues for the Hessian with perturbed parameters at $\lambda = 0$, where the exact ground state is known.
We compute $L_H$ using 5 randomly sampled sets of perturbed parameters.
Furthermore, Figure~\ref{fig:S2} (b) shows how $L_H$ changes by varying the size of the uniform distribution for the perturbation of the circuit parameters from their optimized values.
For the numerical plots, we use the adjusted formula for $L_H$ and the resolved-eigenvalue protocol below to estimate the size of the convex region, $r_{\mathrm{PL}}$, given by Eq.~\ref{eq:r_pl_1} with the curvature lower bound $2\gamma\deltamin$ of Eq.~\ref{eq:r_pl_2} replaced by its resolved numerical estimate $\zeta_{\rm res}$.
We can see that $L_H$ remains roughly constant for perturbations with magnitudes less than $r_{\mathrm{PL}}$.
Therefore, we choose a heuristic radius $r = 10^{-3}$, for which $\|\delta\bth\|_2\leq\sqrt{M}\,r\approx0.035<r_{\mathrm{PL}}\approx0.039$ at $M=1216$ (Fig.~\ref{fig:S2}~(b)), to sufficiently sample the local neighborhood around the optimized parameters whilst still being within the convex region itself.

\begin{figure}
\centering
\includegraphics[width=0.55\linewidth]{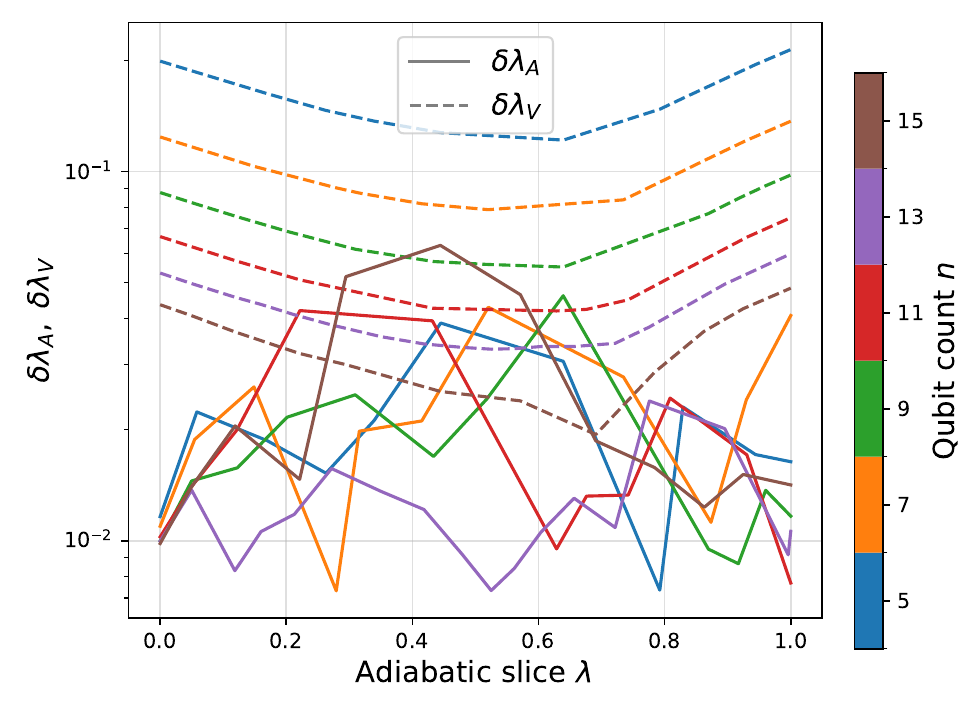}
\caption{\label{fig:S3}
Adiabatic tracking step sizes $\dl_{\mathrm{A}}$ and $\dl_{\mathrm{V}}$ along the adiabatic path for different system sizes.
To optimize simulation run-time, we compute each quantity every five adiabatic slices.
We see that both quantities remain approximately constant along the path, yielding a total of $\sim 100$ slices across all system sizes.
}    
\end{figure}

Numerically, estimating $\dl_{\mathrm{A}}$ requires separating gauge modes from physical Hessian modes.
At an approximate minimum, there is no threshold-independent definition of a smallest nonvanishing eigenvalue.
Near-zero gauge-associated eigenvalues may have either sign away from an exact stationary point.
We therefore define \(\zeta_{\rm res}\) by the following heuristic protocol, using the spectrum at \(\lambda=0\) as a reference.
Note that the size of the Hessian is given by the number of parameters, and remains relatively small by diagonalization standards for all simulations shown in this manuscript, with a maximum size of $(\num{3776} \times \num{3776})$ for $15$ qubits and $64$ layers using the ansatz shown in Fig.~\ref{fig:S1}.
The protocol for choosing \(\zeta_{\rm res}\) is:
\begin{enumerate}
\item Symmetrize the numerical Hessian, compute its signed eigenvalues at point $\lambda$, and sort them by magnitude.
\item Take the ratio between the magnitudes of nearest neighbors in the sorted list.
\item Identify drops in magnitude where the ratio is more than five standard deviations away from the mean ratio.
\item
\begin{itemize}
\item For $\lambda = 0$: the exact ground state is known, so the drop to machine precision separates the gauge modes from the physical spectrum.
Define $\zeta_{\mathrm{tol}}$ as the smallest eigenvalue magnitude above this drop, and define $\zeta_{\rm res}$ at $\lambda=0$ as this same eigenvalue.
\item For $\lambda \neq 0$: define the physical cluster as the eigenvalues of magnitude at least $\zeta_{\mathrm{tol}}$ lying above the largest detected drop, and define $\zeta_{\rm res}$ as the smallest positive eigenvalue in this cluster.
\end{itemize}
\end{enumerate}
This protocol is an empirical diagnostic rather than a proof of the nondegenerate-geometry assumption.
It gives a stable numerical estimate in our data and keeps $\dl$ approximately constant throughout the adiabatic procedure (see Fig.~\ref{fig:S3}).

\begin{figure}
\centering
\begin{overpic}[width=\linewidth]{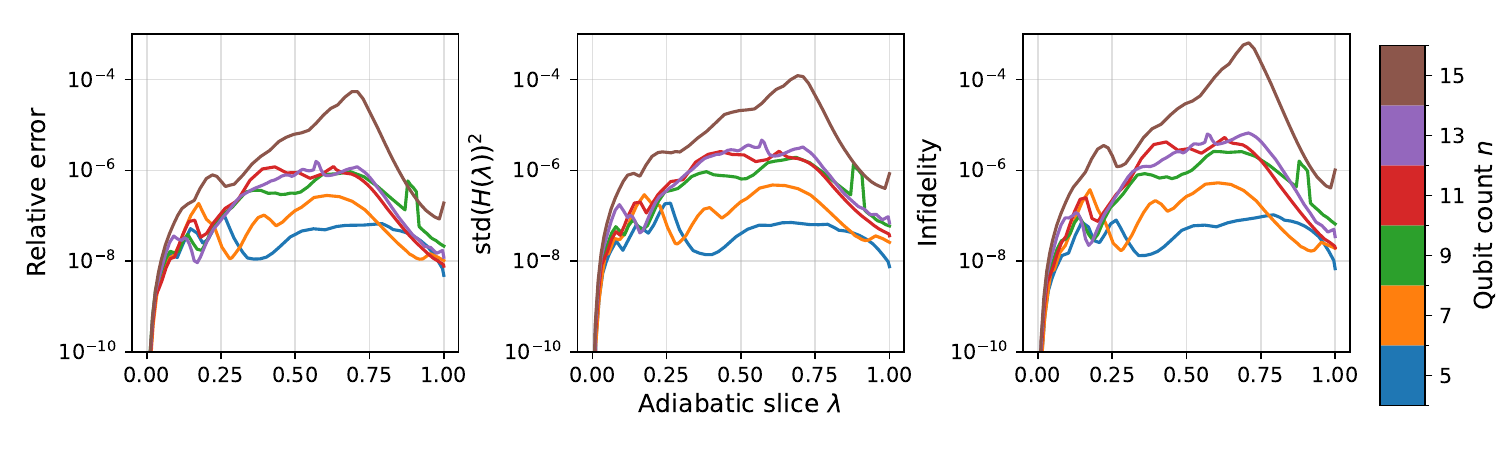}
\put(1,27){(a)}
\put(31,27){(b)}
\put(61,27){(c)}
\end{overpic}
\caption{\label{fig:S4}
Relative error, squared standard deviation of the energy (energy variance) as a function of the adiabatic parameter $\lambda$, and infidelity between the optimized variational circuit and the exact ground state.
We observe that all three measures follow the same behavior for all system sizes.
}
\end{figure}

In Fig.~\ref{fig:S4}, we show that the protocol follows the correct ground state and that the energy variance is a good indicator of the relative energy error as well as the infidelity with respect to the true ground state.

\begin{figure}
\centering
\begin{overpic}[width=\linewidth]{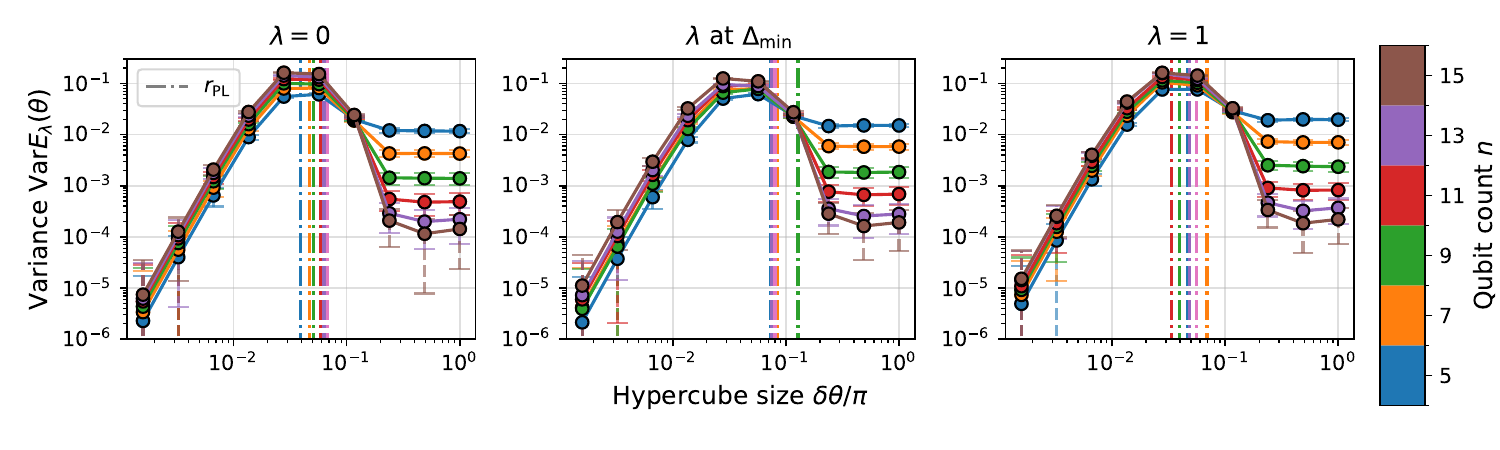}
\put(6,27){(a)}
\put(35,27){(b)}
\put(64,27){(c)}
\end{overpic}
\caption{\label{fig:S5}
Energy variance w.r.t.\ the hypercube size $\delta\theta/\pi$ for multiple points along the adiabatic path.
At the sampled points, the behavior of the variance remains qualitatively consistent.
We also plot $r_{\rm PL}$ (vertical dot-dashed lines), i.e.\ the approximate size of the convex region, evaluated at the optimal parameters.
}
\end{figure}

Finally, Fig.~\ref{fig:S5} shows additional energy-variance data at several points along the adiabatic path.

\section{Measurement cost and stability under shot noise}
\label{sec:measurement_cost}

In realistic implementations, gradients of the variational energy are not evaluated precisely but are estimated from a finite number of projective measurements on quantum hardware.
In this section, we quantify how the required measurement (shot) budget must scale with problem parameters to guarantee, with high probability, that (i) all optimization iterates remain inside the local PŁ region used in the adiabatic tracking analysis, and (ii) the iterates converge toward the \emph{correct} instantaneous variational minimizer along the adiabatic path.
In addition, we derive a one-sided upper confidence bound for the measured standard deviations, so that the run-time verification remains conservative under finite sampling.

Throughout this section, \(M\) denotes the number of variational parameters and \(\nh\) denotes the number of Pauli measurement terms (or commuting measurement groups) in the Hamiltonian decomposition. We use the explicit PŁ radius established in previous sections,
\begin{equation}
r_{\mathrm{PL}} = \frac{\gamma\deltamin}{L_H}, \qquad L_H = 8\|H\|_{\mathrm{op}} M^{3/2}.
\label{eq:rpl_def}
\end{equation}

\subsection{Stochastic gradient model induced by finite shot count}

Fix an adiabatic parameter \(\lambda\) and consider the variational energy
\[
E_\lambda(\bth) = \langle \psi(\bth) | H(\lambda) | \psi(\bth) \rangle, \qquad H(\lambda)=\sum_{\ell=1}^{\nh} h_\ell(\lambda) P_\ell ,
\]
where \(P_\ell\) are Pauli operators.
Using the parameter-shift rule, gradient descent updates take the stochastic form
\begin{equation}
\bth^{(k+1)} = \bth^{(k)} - \eta \bigl( \nabla_\bth E_\lambda(\bth^{(k)}) + \xi_k \bigr),
\label{eq:sgd_update_shots}
\end{equation}
where \(\xi_k\) denotes the gradient estimation error induced by finite measurement shots.
We assume
\begin{equation}
\mathbb{E}[\xi_k \mid \bth^{(k)}] = 0, \qquad \mathbb{E}\!\left[ \|\xi_k\|_2^2 \mid \bth^{(k)} \right] \le \sigma^2(S),
\label{eq:noise_model}
\end{equation}
where \(S\) is the number of shots per Hamiltonian term.
For Pauli measurements with independent shot noise, we can derive an explicit bound on the variance of a single energy estimate.
Fix \(\lambda\) and \(\bth\) and write
\[
E_\lambda(\bth)=\sum_{\ell=1}^{\nh} h_\ell(\lambda)\,\langle P_\ell\rangle_{\bth,\lambda},
\]
where \(\langle P_\ell\rangle_{\bth,\lambda}=\langle\psi(\bth)|P_\ell|\psi(\bth)\rangle\in[-1,1]\).
Assume we estimate each Pauli expectation value \(\langle P_\ell\rangle_{\bth,\lambda}\) by performing \(S\) shots of the measurement of \(P_\ell\), yielding outcomes \(X_{\ell,1},\dots,X_{\ell,S}\in\{+1,-1\}\) with
\[
\mathbb{E}[X_{\ell,s}]=\langle P_\ell\rangle_{\bth,\lambda}, \qquad \mathrm{Var}(X_{\ell,s}) = 1-\langle P_\ell\rangle_{\bth,\lambda}^2 \le 1 .
\]
Define the sample mean estimator
\[
\widehat{\langle P_\ell\rangle} =\frac{1}{S}\sum_{s=1}^S X_{\ell,s}, \qquad \widehat{E_\lambda(\bth)}=\sum_{\ell=1}^{\nh} h_\ell(\lambda)\,\widehat{\langle P_\ell\rangle}.
\]
Then \(\widehat{E_\lambda(\bth)}\) is unbiased:
\[
\mathbb{E}\!\left[\widehat{E_\lambda(\bth)}\right] =\sum_{\ell=1}^{\nh} h_\ell(\lambda)\,\mathbb{E}\!\left[\widehat{\langle P_\ell\rangle}\right] =\sum_{\ell=1}^{\nh} h_\ell(\lambda)\,\langle P_\ell\rangle_{\bth,\lambda} =E_\lambda(\bth).
\]
To bound the variance, first note that, by independence of shots for a fixed \(\ell\),
\[
\mathrm{Var}\!\left(\widehat{\langle P_\ell\rangle}\right) = \mathrm{Var}\!\left(\frac{1}{S}\sum_{s=1}^S X_{\ell,s}\right) = \frac{1}{S^2}\sum_{s=1}^S \mathrm{Var}(X_{\ell,s}) = \frac{1}{S}\mathrm{Var}(X_{\ell,1}) = \frac{1-\langle P_\ell\rangle_{\bth,\lambda}^2}{S} \le \frac{1}{S}.
\]
Next, assume that the estimators \(\widehat{\langle P_\ell\rangle}\) are obtained from independent measurement data across different \(\ell\) (e.g.\ by allocating disjoint shot batches per term).
Then the random variables \(\widehat{\langle P_\ell\rangle}\) are independent, and therefore
\begin{align}
\mathrm{Var}\!\left[\widehat{E_\lambda(\bth)}\right]
&= \mathrm{Var}\!\left[\sum_{\ell=1}^\nh h_\ell(\lambda)\,\widehat{\langle P_\ell\rangle}\right] = \sum_{\ell=1}^\nh h_\ell(\lambda)^2\,\mathrm{Var}\!\left(\widehat{\langle P_\ell\rangle}\right) \label{eq:var_sum_indep} \\
&\le \sum_{\ell=1}^\nh h_\ell(\lambda)^2 \cdot \frac{1}{S} = \frac{1}{S}\sum_{\ell=1}^\nh h_\ell(\lambda)^2 = \frac{\|h(\lambda)\|_2^2}{S},
\end{align}
where we defined
\[
\|h(\lambda)\|_2^2 = \sum_{\ell=1}^\nh h_\ell(\lambda)^2.
\]
If measurements are grouped into commuting sets, the same derivation holds with \(\ell\) indexing the groups, \(h_\ell(\lambda)P_\ell\) replaced by the corresponding grouped observable \(G_\ell(\lambda)\), and \(\|h(\lambda)\|_2^2\) replaced by \(\sum_\ell\|G_\ell(\lambda)\|_{\mathrm{op}}^2\).

A full parameter-shift gradient requires \(2M\) energy evaluations.
Summing the variances of the gradient components yields
\begin{equation}
\sigma^2(S) \le \frac{2M \, \sup_{\lambda \in [0,1]} \|h(\lambda)\|_2^2}{S}.
\label{eq:sigma_scaling_ps}
\end{equation}

\subsection{High-probability invariance of the PŁ region}

Fix \(\lambda\) and write \(\bth^\star=\bth^\star_\lambda\).
Throughout this section, we work inside the PŁ region (see \lref{lem: PŁ inequality})
\[
\mathcal D_\lambda = \{\bth:\ \|\bth-\bth^\star\|_2\le r_{\mathrm{PL}}\}, \qquad r_{\mathrm{PL}}=\frac{\gamma\deltamin}{L_H}, \qquad \mu=\gamma\deltamin .
\]
Consider the stochastic update
\begin{equation}
\bth^{(k+1)}=\bth^{(k)}-\eta\bigl(\nabla_\bth E_\lambda(\bth^{(k)})+\xi_k\bigr), \qquad \eta\le \frac{1}{L},
\label{eq:sgd_update_recall}
\end{equation}
where $L$ is the smoothness constant defined in \lref{lem: smoothness}.

By the triangle inequality applied to~\eqref{eq:sgd_update_recall},
\begin{align}
\|\bth^{(k+1)}-\bth^\star\|_2
&= \bigl\|\bth^{(k)}-\bth^\star-\eta\nabla_\bth E_\lambda(\bth^{(k)})-\eta\xi_k\bigr\|_2 \notag\\
&\le \underbrace{\|\bth^{(k)}-\bth^\star-\eta\nabla_\bth E_\lambda(\bth^{(k)})\|_2}_{\text{noiseless GD step}} +\eta\|\xi_k\|_2.
\label{eq:inv_triangle}
\end{align}
Inside \(\mathcal D_\lambda\), strong convexity and smoothness imply that gradient descent with \(\eta\le 1/L\) is contractive in distance to \(\bth^\star\) (by the argument of \lref{lem: warm-start}):
\begin{equation}
\|\bth^{(k)}-\bth^\star-\eta\nabla_\bth E_\lambda(\bth^{(k)})\|_2 \le \sqrt{1-\eta\mu}\|\bth^{(k)}-\bth^\star\|_2.
\label{eq:inv_contraction}
\end{equation}
Combining~\eqref{eq:inv_triangle}--\eqref{eq:inv_contraction} and using \(\|\bth^{(k)}-\bth^\star\|_2\le r_{\mathrm{PL}}\) gives
\[
\|\bth^{(k+1)}-\bth^\star\|_2 \le \sqrt{1-\eta\mu}\, r_{\mathrm{PL}}+\eta\|\xi_k\|_2.
\]
Thus \(\|\bth^{(k+1)}-\bth^\star\|_2\le r_{\mathrm{PL}}\) is guaranteed whenever \(\sqrt{1-\eta\mu}\,r_{\mathrm{PL}}+\eta\|\xi_k\|_2< r_{\mathrm{PL}}\).
Since \(1-\sqrt{1-\eta\mu}\geq\eta\mu/2\), it suffices that
\begin{equation}\label{eq:noise_abs_opt}
\|\xi_k\|_2<\frac12 \mu\, r_{\mathrm{PL}}.
\end{equation}

One-step invariance keeps every iterate inside \(\mathcal D_\lambda\), but the slice-by-slice warm-start induction of \sref{sec: pl proof} requires slightly more: the terminal iterate must return to the smaller radius \(r_{\mathrm{PL}}/2\), so that it remains a valid warm start after the parameter step to \(\lambda+\delta\lambda\).
Iterating \eqref{eq:inv_triangle}--\eqref{eq:inv_contraction} and using \(\sum_{k=0}^{K-1}(1-\eta\mu)^{k/2}\le\bigl(1-\sqrt{1-\eta\mu}\bigr)^{-1}\le 2/(\eta\mu)\) gives, for any uniform bound \(\|\xi_k\|_2\le\xi_{\max}\),
\begin{equation}
\|\bth^{(K)}-\bth^\star\|_2 \le (1-\eta\mu)^{K/2}\,\|\bth^{(0)}-\bth^\star\|_2 + \frac{2\xi_{\max}}{\mu}.
\label{eq:terminal_radius_bound}
\end{equation}
With \(\|\bth^{(0)}-\bth^\star\|_2\le r_{\mathrm{PL}}\), the two conditions
\begin{equation}
\xi_{\max}\le\frac18\,\mu r_{\mathrm{PL}},
\qquad
K\ge\frac{2}{\eta\mu}\ln 4,
\label{eq:terminal_radius_conditions}
\end{equation}
give \(\|\bth^{(K)}-\bth^\star\|_2\le r_{\mathrm{PL}}/4+r_{\mathrm{PL}}/4=r_{\mathrm{PL}}/2\), and the induction closes.
The tolerance \(\frac18\mu r_{\mathrm{PL}}\), stronger than \eref{eq:noise_abs_opt}, is the one we enforce below.
The second term of \eref{eq:terminal_radius_bound} is a controlled error floor: finite shots do not imply exact convergence at fixed \(S\).

\subsubsection*{High-probability enforcement via sub-Gaussian norm concentration}

Take \(\sigma^2(S)\) equal to the right-hand side of \eref{eq:sigma_scaling_ps}.
Conditionally on the history up to \(\bth^{(k)}\), the \(M\) components of \(\xi_k\) are independent, since each is estimated from its own shifted circuits with fresh shots.
Each component is a sum of independent centered terms \(\pm h_\ell(\lambda)(X_{\ell,s}-\mathbb E X_{\ell,s})/S\) of range \(2|h_\ell(\lambda)|/S\), so Hoeffding's lemma~\cite{Hoeffding63} gives, for every \(\alpha\in\mathbb R^M\),
\begin{equation}
\mathbb E\!\left[e^{\alpha^\top\xi_k}\;\middle|\;\bth^{(k)}\right]
\le
\exp\!\left(\frac{\sigma^2(S)}{2M}\,\|\alpha\|_2^2\right).
\label{eq:subgaussian_mgf}
\end{equation}
The tail inequality for quadratic forms of sub-Gaussian random vectors~\cite{HsKaZh12}, applied with the identity matrix, then gives for all \(x\ge 0\)
\begin{equation}
\mathbb P\!\left(
\|\xi_k\|_2^2\ge\frac{\sigma^2(S)}{M}\bigl(M+2\sqrt{Mx}+2x\bigr)
\;\middle|\; \bth^{(k)}
\right)
\le e^{-x}.
\label{eq:subgaussian_norm_tail}
\end{equation}
Since \(M+2\sqrt{Mx}+2x\le(\sqrt M+\sqrt{2x})^2\), this yields
\begin{equation}
\mathbb P\!\left(
\|\xi_k\|_2 \ge \sigma(S)\Bigl(1+\sqrt{\tfrac{2x}{M}}\Bigr)
\;\middle|\; \bth^{(k)}
\right)
\le e^{-x}.
\label{eq:xi_norm_tail}
\end{equation}
Set \(x=\log(TK/\delta)\), where \(T\) is an upper bound on the number of adiabatic slices of a run.
Then \eqref{eq:xi_norm_tail} gives
\[
\mathbb P\!\left(
\|\xi_k\|_2 \ge \sigma(S)\Bigl(1+\sqrt{\tfrac{2\log(TK/\delta)}{M}}\Bigr)
\;\middle|\; \bth^{(k)}
\right)
\le \frac{\delta}{TK}.
\]
Applying a union bound over all \(TK\) gradient updates along the adiabatic schedule, we obtain that with probability at least \(1-\delta\),
\begin{equation}
\|\xi_k\|_2 \le \sigma(S)\Bigl(1+\sqrt{\tfrac{2\log(TK/\delta)}{M}}\Bigr)
\qquad\text{for every update of every slice.}
\label{eq:xi_uniform_bound}
\end{equation}
Combining \eqref{eq:xi_uniform_bound} with the tolerance \(\xi_{\max}\le \frac18 \,\mu r_{\mathrm{PL}}\) from \eref{eq:terminal_radius_conditions}, it suffices that
\begin{equation}
\sigma(S)\Bigl(1+\sqrt{\tfrac{2\log(TK/\delta)}{M}}\Bigr)
\le \frac18\,\mu r_{\mathrm{PL}}.
\label{eq:sigma_invariance_correct}
\end{equation}
Equivalently,
\begin{equation}
\sigma(S)
\le
\frac18\,\frac{\mu r_{\mathrm{PL}}}{
1+\sqrt{\tfrac{2\log(TK/\delta)}{M}}
}.
\label{eq:sigma_invariance_correct_solved}
\end{equation}
Using the parameter-shift variance proxy \(\sigma^2(S)\le \dfrac{2M\sup_\lambda\|h(\lambda)\|_2^2}{S}\), a sufficient shot budget is therefore
\begin{equation}
S \ge \frac{128\,M\sup_\lambda\|h(\lambda)\|_2^2}{\mu^2 r_{\mathrm{PL}}^2} \left( 1+\sqrt{\tfrac{2\log(TK/\delta)}{M}} \right)^2.
\label{eq:shots_invariance_correct}
\end{equation}
Substituting \(\mu=\gamma\deltamin\) and \(r_{\mathrm{PL}}=\gamma\deltamin/L_H\) yields the gap scaling
\begin{equation}
\boxed{S\ge \frac{128\,M\sup_\lambda\|h(\lambda)\|_2^2\,L_H^2}{\gamma^4\deltamin^4} \left( 1+\sqrt{\tfrac{2\log(TK/\delta)}{M}} \right)^2}\,.
\label{eq:shots_invariance_gap_scaling_correct}
\end{equation}
In particular, the leading dependence on the spectral gap is \(S\propto \deltamin^{-4}\).

For the output test of \sref{subsec:finite_shot_certificate} to pass with relative margin \(\rho\), the terminal state must also satisfy \(\Std_{\psi(\bth^{(K)})}(H(\lambda))^2<(1-\rho)(\Delta_c/4)^2\).
Since \(\Std_{\psi(\bth)}(H(\lambda))\leq W\sqrt M\,\|\bth-\bth^\star\|_2\), with \(W\) the spectral width of Remark~\ref{rem:adaptive_schedule}, \eref{eq:terminal_radius_bound} guarantees this if, in addition to \eref{eq:terminal_radius_conditions},
\begin{equation}
\xi_{\max}\le\frac{\mu\sqrt{1-\rho}\,\Delta_c}{16W\sqrt M},
\qquad
K>\frac{2}{\eta\mu}\ln\frac{8W\sqrt M\,r_{\mathrm{PL}}}{\sqrt{1-\rho}\,\Delta_c}.
\label{eq:margin_conditions}
\end{equation}
This multiplies the right-hand side of \eref{eq:shots_invariance_correct} by \(\max\{1,4MW^2r_{\mathrm{PL}}^2/((1-\rho)\Delta_c^2)\}\), giving \(S\propto\max\{\deltamin^{-4},\deltamin^{-2}\Delta_c^{-2}\}\) up to logarithmic factors.

\subsection{Finite-shot standard-deviation certification}
\label{subsec:finite_shot_certificate}

The deterministic verification conditions compare the energy standard deviations \(\Std_\psi(H(\lambda))\) and \(\Std_\psi(D)\), with \(D=H_\mathrm{f}-H_\mathrm{i}\) the step observable used above, to thresholds set by the gap.
With finitely many shots, these quantities can only be estimated, and soundness requires that statistical fluctuations never make a state look better than it is.
We therefore construct a measurable quantity \(\overline\sigma_\psi(H(\lambda))\) that upper-bounds \(\Std_\psi(H(\lambda))\) with a controlled failure probability; the construction uses only the Pauli decomposition of a Hermitian observable, so it applies verbatim to \(D\).
Replacing each measured standard deviation by its upper confidence bound then keeps every acceptance conservative.

\subsubsection*{Moment estimators}

Fix \(\lambda\) and abbreviate \(H=H(\lambda)\).
After collecting identical strings and separating the identity terms, write
\begin{align*}
H &= a_0 I+\sum_{\ell=1}^{\nh_H}a_\ell P_\ell ,\\
H^2 &= b_0 I+\sum_{r=1}^{\nh_{H^2}}b_r Q_r ,
\end{align*}
where \(P_\ell\) and \(Q_r\) are nonidentity Hermitian Pauli strings and all coefficients are real.
The second decomposition follows by expanding \(H^2\): anticommuting cross terms cancel, so before equal strings are combined, it has at most \(\nh_H(\nh_H+1)/2\) nonidentity terms.
Both moments are therefore accessible term by term, exactly as the energy in the previous subsections.

Write \( \mu_H=\langle H\rangle_\psi \), \( \nu_H=\langle H^2\rangle_\psi \), and \( V_H=\nu_H-\mu_H^2 \) for the two moments and the variance.
Using disjoint batches, measure each \(P_\ell\) on \(N_H\) copies of \(\ket{\psi}\) and each \(Q_r\) on \(N_{H^2}\) copies, where \(N_H,N_{H^2}\) are positive integers.
Denote the respective outcomes by \(X_{\ell,s},Y_{r,s}\in\{-1,+1\}\), with \(\mathbb E X_{\ell,s}=\langle P_\ell\rangle_\psi\) and \(\mathbb E Y_{r,s}=\langle Q_r\rangle_\psi\).
Define
\begin{align*}
\widehat\mu_H
&=
a_0+\sum_{\ell=1}^{\nh_H}a_\ell\overline X_\ell,
&
\overline X_\ell
&=
\frac{1}{N_H}\sum_{s=1}^{N_H}X_{\ell,s},\\
\widehat\nu_H
&=
b_0+\sum_{r=1}^{\nh_{H^2}}b_r\overline Y_r,
&
\overline Y_r
&=
\frac{1}{N_{H^2}}\sum_{s=1}^{N_{H^2}}Y_{r,s}.
\end{align*}
Both estimators are unbiased, but the plug-in variance \(\widehat\nu_H-\widehat\mu_H^2\) is biased downward because \(\mathbb E(\widehat\mu_H^2)=\mu_H^2+\Var(\widehat\mu_H)\).
More importantly, a point estimate can fall below the truth and carries no one-sided guarantee, so a certificate cannot use it directly: we need explicit confidence radii for both moments.

\subsubsection*{One-sided upper confidence bound}

Consider first \(\widehat\mu_H\).
It is a weighted sum of independent bounded outcomes: each summand \(a_\ell X_{\ell,s}/N_H\) has range \(2|a_\ell|/N_H\), so the squared ranges add up to
\[
\sum_{\ell=1}^{\nh_H}\sum_{s=1}^{N_H}\left(\frac{2|a_\ell|}{N_H}\right)^{\!2}
=\sum_{\ell=1}^{\nh_H}N_H\left(\frac{2|a_\ell|}{N_H}\right)^{\!2}
=\frac{4\|a\|_2^2}{N_H}
=:4c_\mu ,
\]
where the identity coefficient is omitted from the norm.
Weighted Hoeffding concentration~\cite{Hoeffding63} therefore gives, for every \(\varepsilon>0\),
\[
\Pr\!\left[|\widehat\mu_H-\mu_H|>\varepsilon\right]
\leq
2\exp\!\left(-\frac{\varepsilon^2}{2c_\mu}\right),
\]
and identically for \(\widehat\nu_H\) with \(c_\nu=\|b\|_2^2/N_{H^2}\).
(If a coefficient vector vanishes, its estimator is exact and its radius may be set to zero.)
To turn the two tails into explicit failure budgets, choose \(\alpha_\mu,\alpha_\nu\in(0,1)\) with \(\alpha_\mu+\alpha_\nu<1\) and equate the right-hand sides to \(\alpha_\mu\) and \(\alpha_\nu\); the corresponding radii are
\[
\varepsilon_\mu
=
\sqrt{2c_\mu\log\!\frac{2}{\alpha_\mu}},
\qquad
\varepsilon_\nu
=
\sqrt{2c_\nu\log\!\frac{2}{\alpha_\nu}}.
\]
A union bound---which requires no independence between the two estimators---shows that the simultaneous event
\[
\mathcal E_H
=
\left\{
|\widehat\mu_H-\mu_H|\leq\varepsilon_\mu,\quad
|\widehat\nu_H-\nu_H|\leq\varepsilon_\nu
\right\}
\]
holds with probability at least \(1-\alpha_\mu-\alpha_\nu\).

On \(\mathcal E_H\), the second moment is at most \(\nu_H\leq\widehat\nu_H+\varepsilon_\nu\).
For the first moment we need a \emph{lower} bound on \(|\mu_H|\), because \(\mu_H^2\) enters the variance with a minus sign: the true mean lies within \(\varepsilon_\mu\) of \(\widehat\mu_H\), so its magnitude can fall below the observed \(|\widehat\mu_H|\) by at most the radius (the reverse triangle inequality), \(|\mu_H|\geq|\widehat\mu_H|-\varepsilon_\mu\), and since \(|\mu_H|\geq0\) always,
\[
|\mu_H|
\geq
m_\mu
:=
\max\!\left\{|\widehat\mu_H|-\varepsilon_\mu,0\right\}.
\]
Combining the two bounds,
\[
V_H=\nu_H-\mu_H^2
\leq
\widehat\nu_H+\varepsilon_\nu-m_\mu^2
\leq
\overline V_H
:=
\max\!\left\{0,\,\widehat\nu_H+\varepsilon_\nu-m_\mu^2\right\},
\]
so on \(\mathcal E_H\) the measurable quantity \(\overline\sigma_\psi(H)=\sqrt{\overline V_H}\) upper-bounds the true standard deviation.
We record the conclusion.

\begin{proposition}[Finite-shot upper confidence bound]
\label{prop:finite_shot_std}
With the estimators, radii, and \(\overline\sigma_\psi(H)\) constructed above,
\[
\Pr\!\left[
\Std_\psi(H)\leq\overline\sigma_\psi(H)
\right]
\geq
1-\alpha_\mu-\alpha_\nu.
\]
Consequently, for any threshold \(\tau>0\), the probability that \(\overline\sigma_\psi(H)\leq\tau\) while \(\Std_\psi(H)>\tau\) is at most \(\alpha_\mu+\alpha_\nu\).
\end{proposition}

The first claim was derived above; for the second, note that \(\overline\sigma_\psi(H)\leq\tau\) together with \(\Std_\psi(H)>\tau\) is impossible on \(\mathcal E_H\), so a false acceptance requires the failure of \(\mathcal E_H\).

\subsubsection*{Shot count}

Proposition~\ref{prop:finite_shot_std} guarantees soundness for \emph{any} shot counts: too few shots never raise the false-acceptance probability above \(\alpha_\mu+\alpha_\nu\), because the radii \(\varepsilon_\mu,\varepsilon_\nu\) only inflate \(\overline\sigma_\psi(H)\).
The price of too few shots is that the test becomes uninformative---\(\overline\sigma_\psi(H)\) may then exceed the threshold even for a perfectly good state.
The remaining question is:
How many shots guarantee that a state whose true variance passes the test with some margin is indeed certified?

The first step is to bound how far \(\overline V_H\) can overshoot the true variance. On the event \(\mathcal E_H\), the second-moment part obeys \(\widehat\nu_H+\varepsilon_\nu\leq\nu_H+2\varepsilon_\nu\): one radius from the fluctuation of \(\widehat\nu_H\), a second added by construction.
For the subtracted term, \(m_\mu\) can undershoot \(|\mu_H|\) by at most two radii,
\[
0\leq|\mu_H|-m_\mu\leq2\varepsilon_\mu
\]
(one radius from the fluctuation of \(\widehat\mu_H\), a second from the subtraction in \(m_\mu\)), and the resulting deficit of the square is controlled by the difference of squares: with \(m_\mu\leq|\mu_H|\leq\|H\|_{\mathrm{op}}\),
\[
\mu_H^2-m_\mu^2
=\bigl(|\mu_H|-m_\mu\bigr)\bigl(|\mu_H|+m_\mu\bigr)
\leq4\|H\|_{\mathrm{op}}\varepsilon_\mu .
\]
Together,
\[
\overline V_H
\leq
V_H+2\varepsilon_\nu+4\|H\|_{\mathrm{op}}\varepsilon_\mu
\]
on \(\mathcal E_H\) (the right-hand side is nonnegative, so the outer \(\max\{0,\cdot\}\) in \(\overline V_H\) preserves the bound).

Now suppose the state passes the test \(\Std_\psi(H)<\tau\) with a relative margin \(\rho\in(0,1)\), i.e.\ \(V_H<(1-\rho)\tau^2\).
By the overshoot bound, \(\overline\sigma_\psi(H)<\tau\) is guaranteed on \(\mathcal E_H\) as soon as the total overshoot stays below the margin, \(2\varepsilon_\nu+4\|H\|_{\mathrm{op}}\varepsilon_\mu\leq\rho\tau^2\).
Splitting this budget evenly between the two radii gives the sufficient conditions \(\varepsilon_\nu\leq\rho\tau^2/4\) and \(\varepsilon_\mu\leq\rho\tau^2/(8\|H\|_{\mathrm{op}})\).

Finally, the radii are explicit functions of the shot counts, \(\varepsilon_\mu^2=2(\|a\|_2^2/N_H)\log(2/\alpha_\mu)\) and \(\varepsilon_\nu^2=2(\|b\|_2^2/N_{H^2})\log(2/\alpha_\nu)\), so the two conditions invert to the sufficient positive integer shot counts
\begin{align}
N_H
&\geq
\frac{128\|a\|_2^2\|H\|_{\mathrm{op}}^2}{\rho^2\tau^4}
\log\!\frac{2}{\alpha_\mu},
\label{eq:resolved_first_moment_shots}\\
N_{H^2}
&\geq
\frac{32\|b\|_2^2}{\rho^2\tau^4}
\log\!\frac{2}{\alpha_\nu}.
\label{eq:resolved_second_moment_shots}
\end{align}
Integer ceilings are understood.
Under the stated variance margin, these counts give \(\Pr[\overline\sigma_\psi(H)<\tau]\geq1-\alpha_\mu-\alpha_\nu\).
For \(\tau=\Theta(\Delta_c)\), the shots per measured Pauli term therefore scale as
\begin{equation}
N_H,N_{H^2}
=
O\!\left(
\Delta_c^{-4}\log\frac{1}{\alpha}
\right),
\label{eq:certificate_gap_scaling}
\end{equation}
where \(\alpha=\min\{\alpha_\mu,\alpha_\nu\}\), with the coefficient norms, \(\|H\|_{\mathrm{op}}\), and \(\rho\) fixed.
Here \(N_H\) and \(N_{H^2}\) count the shots spent on each individual Pauli string; a full variance check measures \(\nh_H+\nh_{H^2}\) strings, so its total cost carries the term counts as additional factors, see \eref{eq:single_certificate_cost} below.
In \eref{eq:resolved_first_moment_shots}, \(\|H\|_{\mathrm{op}}\) may be replaced by any known upper bound---e.g.\ the coefficient sum \(|a_0|+\sum_\ell|a_\ell|\), since every Pauli string has unit operator norm; this only increases the prescribed shot count.

Nothing above uses more than the Pauli decompositions of a Hermitian observable and its square.
The construction therefore applies verbatim to the step observable \(D\), yielding an upper confidence bound \(\overline\sigma_\psi(D)\) with its own coefficient vectors, shot counts \(N_D\) and \(N_{D^2}\), and failure budgets.
(In the degenerate case \(D=0\), every estimator is exact, and no shots are required.)

\subsubsection*{Certificate calls along the adiabatic chain}

Proposition~\ref{prop:finite_shot_std} concerns a single evaluation of one observable on one state.
A full run makes at most \(2T+1\) such calls: one at \(\lambda=0\) and, at each of at most \(T\) slices, one for \(H(\lambda_t)\) and one for \(D\).
Assign every call the failure budgets
\[
\alpha_\mu=\alpha_\nu=\frac{\delta_{\rm cert}}{2(2T+1)},
\qquad
\delta_{\rm cert}\in(0,1),
\]
so that the budgets of all calls sum to at most \(\delta_{\rm cert}\).
The calls are adaptive---the state tested at slice \(t\) depends on all previous measurement outcomes---but this is harmless: the state, decomposition, shot counts, and budgets of each call are fixed before its fresh shots are drawn, so the Hoeffding tails behind Proposition~\ref{prop:finite_shot_std} hold conditionally on the entire previous history, and a union bound over these calls needs no independence.
The bound \(T\) is known in advance, because every step of a run except the last is at least \(\min\{\delta\lambda_{\rm A},\kappa\Delta_c/(4\sum_\ell|d_\ell|)\}\), with \(\delta\lambda_{\rm A}\) the step cap of the run and \(d_\ell\) the nonidentity Pauli coefficients of \(D\): after every accepted slice, including \(\lambda=0\), the numerator of \eref{eq:finite_shot_step_proposal} is at least \(\Delta_c/4\), and since \(\Std_\psi(D)\leq\sum_\ell|d_\ell|\) for every state (Popoviciu's inequality), \(\overline\sigma_\psi(D)\) is replaced by \(\min\{\overline\sigma_\psi(D),\sum_\ell|d_\ell|\}\).

Consequently, with probability at least \(1-\delta_{\rm cert}\), every confidence event along the chain holds simultaneously.
On this event, every \(\overline\sigma\) issued during the run is a valid upper bound; and if, in addition, the optimizer delivers at every slice a state that passes the acceptance test with relative margin \(\rho\), \(V_{H}<(1-\rho)(\Delta_c/4)^2\), while the shot counts obey \eref{eq:resolved_first_moment_shots}--\eref{eq:resolved_second_moment_shots}, then every standard-deviation test accepts.
In other words, the probability that the chain produces even one invalid certificate, or, under these conditions, one such rejection, over all \(\lambda\) slices, is at most \(\delta_{\rm cert}\).

A rejection, therefore, signals this rare statistical event, a genuine failure of the optimizer, or too few shots.
In any case, the run ends and the protocol restarts as in Remark~\ref{rem:adaptive_schedule} (line~12 of Algorithm~1 of the main text), with the per-run failure budgets of Remark~\ref{rem:adaptive_schedule}.
A rejection asserts nothing, so soundness is never at stake.

\subsubsection*{Application to run-time verification}

At slice \(t\), apply Proposition~\ref{prop:finite_shot_std} at \(\lambda=\lambda_t\) and accept the optimized output (at \(t=0\), the initial state) only if
\begin{equation}
\overline\sigma_{\psi_t}(H(\lambda_t))
\leq
\frac{\Delta_c}{4}.
\label{eq:finite_shot_runtime_acceptance}
\end{equation}
On the simultaneous confidence event, \(\Std_{\psi_t}(H(\lambda_t))\leq\overline\sigma_{\psi_t}(H(\lambda_t))\) and \(\Std_{\psi_t}(D)\leq\overline\sigma_{\psi_t}(D)\), so acceptance implies the exact condition \( \Std_{\psi_t}(H(\lambda_t))\leq\Delta_c/4 \).
The finite-shot step proposal is
\begin{equation}
\overline{\delta\lambda}_{\rm V}
=
\kappa
\frac{
\Delta_c/2-\overline\sigma_{\psi_t}(H(\lambda_t))
}{
\overline\sigma_{\psi_t}(D)
},
\qquad 0<\kappa<1.
\label{eq:finite_shot_step_proposal}
\end{equation}
If \(\overline\sigma_{\psi_t}(D)=0\), the confidence event also gives \(\Std_{\psi_t}(D)=0\), so the propagation bound imposes no verification cap.
When \(\Std_{\psi_t}(D)>0\), the proposal obeys
\[
\overline{\delta\lambda}_{\rm V}
\leq\kappa
\frac{\Delta_c/2-\Std_{\psi_t}(H(\lambda_t))}
{\Std_{\psi_t}(D)}
<\delta\lambda_{\rm V}.
\]
Hence choosing \(|\delta\lambda|\leq\overline{\delta\lambda}_{\rm V}\) cannot violate the exact step condition.

With finite-shot gradients, the energy can increase, while the proof of item (4) of Theorem~\ref{thm:verification_pl_tracking} uses \(\mu_{\rm out}\le\mu_{\rm in}\).
With finite shots, an output at slice \(t+1\) is therefore accepted only if also
\begin{equation}
\widehat\mu_{H_{t+1}}+\varepsilon_{\mu,H_{t+1}}
\leq
\widehat\mu_{H_{t}}+\delta\lambda\,\widehat\mu_{D,t}
-\varepsilon_{\mu,H_{t}}-\delta\lambda\,\varepsilon_{\mu,D,t}
+\frac{\Delta_c}{4},
\label{eq:finite_shot_energy_check}
\end{equation}
with the first-moment estimates and radii of the certificate calls on \(\ket{\psi_t}\) and \(\ket{\psi_{t+1}}\).
On the confidence events this gives \(\mu_{\rm out}\le\mu_{\rm in}+\Delta_c/4\), which suffices there, since \(\mu_{\rm in}<E_0+\Delta_c/2\) and an excited branch would require \(\mu_{\rm out}\ge E_0+3\Delta_c/4\).
No additional measurements are needed.
On the confidence events of runs that satisfy \eref{eq:margin_conditions}, the test also passes: since \((H-E_0)\ket{\psi_0}=0\), \(\mu_{\rm out}-E_0=\langle\psi-\psi_0|(H-E_0)|\psi-\psi_0\rangle\leq WM\|\bth_{\rm out}-\bth^\star\|_2^2\leq\Delta_c/16\) and \(\mu_{\rm in}\geq E_0\), while \eref{eq:resolved_first_moment_shots} gives \(\varepsilon_{\mu,H}\leq\Delta_c/64\), so it suffices to choose \(N_D\geq512\|d\|_2^2\log(2/\alpha_\mu)/\Delta_c^2\), which gives \(\varepsilon_{\mu,D,t}\leq\Delta_c/16\).

With the allocation above, the implications of this statement, \eref{eq:finite_shot_runtime_acceptance}, and \eref{eq:finite_shot_energy_check} hold at every call with probability at least \(1-\delta_{\rm cert}\), so every accepted output is certified; together with \eref{eq:xi_uniform_bound} and the step, iteration, and shot conditions above, all tests of a run also pass with probability at least \(1-\delta-\delta_{\rm cert}\).

\subsection{Total measurement cost}

Each parameter-shift gradient evaluation requires \(2M\) energy measurements.
Each energy measurement consists of \(\nh\) Pauli terms estimated with \(S\) shots.
Therefore, the measurement cost per gradient step is
\[
\mathcal{C}_{\mathrm{iter}} = 2\,M\,\nh \,S .
\]
Over \(K\) gradient updates, the total measurement cost (for fixed $\lambda$) is
\begin{equation}
\mathcal{C}_{\mathrm{total}} = 2\,K\,M\,\nh\, S ,
\end{equation}
with \(S\) chosen according to~\eqref{eq:shots_invariance_gap_scaling_correct}, multiplied for certification by the factor following \eref{eq:margin_conditions}.

One standard-deviation certificate of \(H_t=H(\lambda_t)\) or of \(D\) has the ungrouped cost
\begin{equation}
\mathcal C_{\rm cert}(H_t)
=
\nh_{H_t} N_{H_t}+\nh_{H_t^2}N_{H_t^2},
\qquad
\mathcal C_{\rm cert}(D)
=
\nh_{D} N_{D}+\nh_{D^2}N_{D^2},
\label{eq:single_certificate_cost}
\end{equation}
where the per-term shot counts are chosen according to \eref{eq:resolved_first_moment_shots}--\eref{eq:resolved_second_moment_shots}, with \(N_D\) also satisfying the condition below \eref{eq:finite_shot_energy_check}.
The second-moment contributions are essential: estimating \(\langle H\rangle\) or \(\langle D\rangle\) alone does not estimate the corresponding variance.

The algorithm evaluates both standard deviations once per accepted slice.
Consequently, the ungrouped circuit-execution count for such a slice is
\begin{align}
\mathcal C_{\rm slice}
={}&
2KM \nh S
+\nh_{H_t} N_{H_t}+\nh_{H_t^2}N_{H_t^2}
\notag\\
&+
\nh_DN_D+\nh_{D^2}N_{D^2}.
\label{eq:total_measurement_cost_per_slice}
\end{align}
A restarted run incurs the cost \eref{eq:total_measurement_cost_per_slice} of its slices again; by Remark~\ref{rem:adaptive_schedule}, all runs together cost at most twice the a priori cost bound of the final run.
Over the full path, the total cost is the sum of \eref{eq:total_measurement_cost_per_slice} over all attempted slices, plus one certificate of \(H(0)\) per run.

The gap dependence \(O(\Delta_c^{-4})\) in \eref{eq:certificate_gap_scaling} applies to the shots per measured term for a fixed relative variance margin.
The total per-slice cost is the full expression \eref{eq:total_measurement_cost_per_slice}.

\subsubsection*{Summary}

When the PŁ radius is fixed to its explicit value \(r_{\mathrm{PL}}=\gamma\deltamin/L_H\), finite-shot noise does not invalidate adiabatic PŁ-based tracking, provided the shot budget scales polynomially in \(M\) and the inverse gap and logarithmically in \(K\).
With parameter-shift gradients, both high-probability invariance of the PŁ region and convergence to a controlled neighborhood of the correct instantaneous minimizer are guaranteed under an explicit and controlled measurement budget.
The run-time certificate remains conservative under finite sampling when each measured standard deviation is replaced by the one-sided upper confidence bound of \sref{subsec:finite_shot_certificate}.

\bibliography{references}